\documentclass[superscriptaddress,twocolumn,amssymb,nofootinbib,10pt]{revtex4-2}
\usepackage[normalem]{ulem}
\usepackage[utf8]{inputenc}
\usepackage{comment}
\usepackage{graphicx}
\usepackage{graphics}
\usepackage{amsmath, amssymb}
\usepackage{graphics,bm}
\usepackage{graphicx}
\usepackage{bbold}
\usepackage{slashed}
\usepackage{feynmf}
\usepackage{wrapfig}
\usepackage{hyperref}
\usepackage{cancel}
\usepackage[usenames]{color}  

\usepackage[compat=1.1.0]{tikz-feynman}
\usepackage{tikz}
\usetikzlibrary{arrows,shapes}
\usetikzlibrary{trees}
\usetikzlibrary{matrix,arrows} 			
\usetikzlibrary{positioning}				
\usetikzlibrary{calc,through}				
\usepackage{pgffor}							

\newcommand{\be}{\begin{equation}}
\newcommand{\ee}{\end{equation}}
\newcommand{\beq}{\begin{equation}}
\newcommand{\eeq}{\end{equation}}

\newcommand{\bqa}{\begin{eqnarray}}
\newcommand{\eqa}{\end{eqnarray}}

\newcommand{\ms}{\overline{\text{\tiny MS}}}

\def\square{\vcenter{\vbox{\hrule height.4pt
          \hbox{\vrule width.4pt height4pt
          \kern4pt\vrule width.3pt}\hrule height.4pt}}}

\usepackage[normalem]{ulem}

\begin{document}

\title{Color superconductivity and speed of sound in the two-flavor quark-meson diquark model}

\author{Jens O. Andersen}
\email{jens.andersen@ntnu.no}
\affiliation{Department of Physics, Faculty of Natural Sciences,NTNU, 
Norwegian University of Science and Technology, H{\o}gskoleringen 5,
N-7491 Trondheim, Norway}

\author{Mathias P. N{\o}dtvedt}
\email{mathias.p.nodtvedt@ntnu.no}

\affiliation{Department of Physics, Faculty of Natural Sciences,NTNU, 
Norwegian University of Science and Technology, H{\o}gskoleringen 5,
N-7491 Trondheim, Norway}

\date{\today}

\begin{abstract}
We discuss the properties of the two-flavor quark-meson diquark (QMD)
model as a renormalizable low-energy model for QCD in the 2SC phase of QCD. The effective degrees of freedom are the mesons (sigma and pions), quarks, and diquarks. Some of the parameters of the model can be determined by expressing them in terms of
the vacuum meson masses and the pion decay constant using the on-shell renormalization scheme. The remaining parameters are considered free, although they in principle can be calculated
from QCD. The thermodynamic potential is calculated in a mean-field approximation taking only quark loops into account.  In this approximation, we derive a set of renormalization group equations for the running masses and couplings. 
The solutions to these equations are used to  improve the thermodynamic potential $\Omega$ and thereby thermodynamic quantities. Four parameter sets are chosen and the phase diagram in the $\bar{\mu}$--$T$  plane is obtained (with $\bar{\mu}={1\over3}\mu_B$). We also
calculate the speed of sound $c_s$ 
as a function of $\bar{\mu}$ at vanishing temperature.
For large values of $\bar{\mu}$, the speed of sound approaches the conformal limit $c_s={1\over\sqrt{3}}$ from above, in disagreement with perturbative calculations, but agreement with hard-dense-loop resummed perturbation theory.
\end{abstract}

\maketitle

\section{Introduction}
\begin{figure}[htb!]
\includegraphics[width=9cm]{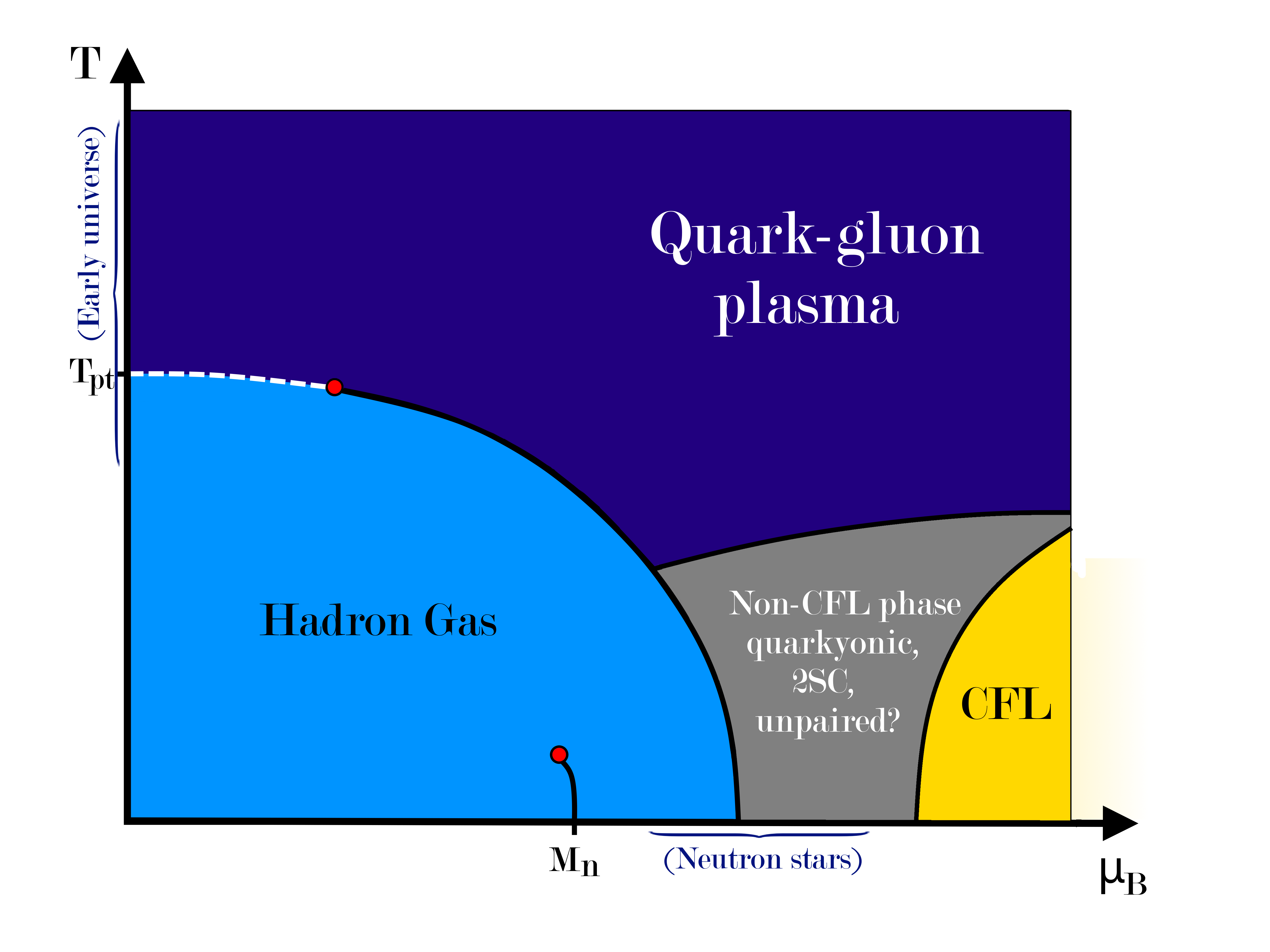}
\caption{Phase diagram of QCD in the ${\mu_B}$--$T$ plane.
See main text for details.}%
\label{muit}
\end{figure}

The phase diagram of quantum chromodynamics (QCD) has received a lot of attention in recent decades due its relevance for the early universe, heavy-ion collisions, and compact 
stars~\cite{raja,alford,fukurev}. The phases of QCD are conventionally shown in a diagram spanned by the ${\mu}_B$ and $T$-axes, see Fig.~\ref{muit}. 

For $\mu_B=0$, lattice simulations have been used to calculate with high precision the thermodynamics  properties of QCD and the crossover temperature for the transition between the hadronic
phase and the quark-gluon plasma phase~\cite{lat1,lat2}. At finite baryon chemical potential, lattice simulations are
hampered by the sign problem. The fermion determinant is complex
for $\mu_B\neq0$ such that standard importance sampling techniques cannot be used.
However, for small values of $\mu_B$, one can enter the $\mu_B$--$T$ plane from the $T$-axis
using Taylor expansions about $\mu_B=0$. The situation is very different in the case of nonzero isospin chemical potential $\mu_I$ or a nonzero magnetic background $B$
(both with $\mu_B=0$). The functional determinant is manifestly real and lattice simulations can be performed. In recent years, high-precision Monte Carlo simulations have been carried out
for nonzero $B$ and nonzero $\mu_I$. In the former case,
focus has been on the chiral transition temperature
as a function of $B$, in particular the mechanism behind
inverse magnetic catalysis. In the latter case,
it has been the critical temperature for pion condensation as a function of $\mu_I$, the equation of state (EoS), and the speed of sound. Lattice simulations have been supplemented by work using  low-energy  effective theories (chiral perturbation theory) and models (NJL and quark-meson models and their Polyakov-loop extended counterparts).
See e.g. Refs.~\cite{revhon,fariasrev,gergyrev,longpaper,endrodieos} for reviews.

The sign problem makes it prohibitively difficult to calculate thermodynamic quantities of
QCD at 
low temperature and large $\mu_B$ using first-principle methods, exactly the region where it is relevant for compact stars. The inapplicability of lattice simulations forces one to use other methods. 
One of these methods is perturbative QCD (pQCD). Perturbative QCD can be used when the strong coupling
$\alpha_s$ is small. At low temperature, this requires large baryon densities or equivalently large baryon chemical potentials. Weak-coupling expansions in cold dense QCD has a long history dating back to the 1970s~\cite{freed1,freed2,baluni}. However, such calculations are complicated by the fact that there are two different momentum scales,
the hard scale $\mu_q$ (with $\mu_q$ being a quark chemical potential) and the soft scale
$m_E\simeq\alpha_{s}^{1/2}\mu_q$, where $\alpha_s$ is the strong coupling constant.
For gluons, the situation is more complicated. For hard gluons, one can use the naive
loop expansion or naive perturbation theory. Soft gluons, on the other hand, require
resummation, which is done by using the hard-dense-loop approximation~\footnote{Even at $T=0$, this is often referred to as hard thermal loops.}.
In recent years, there has been significant progress in perturbative QCD calculations 
of the thermodynamic quantities in cold dense QCD~\cite{finns1,finns2,finns20,finns3}. The perturbative expansion for massless quarks is now known to order $\alpha_s^3\log\alpha_s$.~\footnote{An approach based on the renormalization group has been used to sum up leading soft logarithms of the cold dense pressure to all orders in $\alpha_s$, showing reduced renormalization scale dependence~\cite{loic}.}
It is estimated that the expansions for the thermodynamic quantities converge at above 25-40 times the nuclear saturation density. 
An important question is whether the perturbative expansion of the EoS
at very high densities can be used to constrain the EoS at densities found in compact stars by matching them. The impact of pQCD seems to depend heavily on the matching density chosen,
see e.g. Refs.~\cite{evidence0,evidence1} for a discussion.

The perturbative expansion is an expansion around the trivial vacuum i.e. with no  fermion pairing. This is not the correct ground state at high density. Due to an attractive interaction channel provided by one-gluon exchange, the Fermi surface
is rendered unstable, which leads to the formation of Cooper pairs~\cite{frau,barrois,bailin}.
At large densities, where one can ignore the masses of the quarks, quarks with all colors and flavors participate on an equal footing in the formation of zero-momentum and zero spin diquark pairs~\cite{alfordcolor,rappcolor,alfordcolor2}. This leads to the so-called color-flavor locked (CFL) phase of QCD at the highest densities. In this phase, the symmetry-breaking pattern is $U(1)_B\times SU(3)_L\times SU(3)_R\times SU(3)_c\rightarrow SU(3)_{L+R+c}$.
While the symmetry-breaking pattern is similar to that in the vacuum, the mechanism is very different. In the vacuum, the appearance of a quark condensates mixes left-handed and
right-handed fields. In the CFL phase, the left-handed and right-handed condensates are separately invariant under the subgroups SU(3)$_{L+c}$ and SU(3)$_{R+c}$, respectively. The color-flavor orientations of the condensates are locked  together by the color
transformations. As in the vacuum, there are eight pseudo Goldstone bosons associated with the breaking of the nonabelian symmetries, and a single Goldstone boson associated with breaking of the U(1)$_B$ baryon symmetry.

For lower values of the baryon chemical potential $\mu_B$, one can no longer ignore  the mass of the $s$-quark. This gives rise to stress on the CFL phase. Generally, Cooper pairing between two species is suppressed if their Fermi momenta are 
sufficiently different. This is what happens here since the Fermi momenta of the $s$-quark is smaller than that of the $d$-quark. Similarly, the neutrality constraint, gives rise to 
a Fermi momentum for the $u$-quark which is different from that of the other quark flavors. For relatively small stress, it is still advantageous to be in the CFL phase, albeit a modified
version involving a rotation in flavor space~\cite{bedak}. This rotation may be interpreted as condensation of neutral kaons, which takes place when the relevant chemical potential is
larger than the mass of the condensing meson. For even lower values of $\mu_B$, the stress on the pairing due to the mass of the $s$-quark is so large that the CFL phase is no longer favorable. The quarkyonic phase~\cite{kojo,gloss}
is then a possibility as is the so-called 2SC phase.
In the latter phase, the massless $u$ and $d$-quarks with two out of three colors (e.g. red and green) pair up such that the diquark condensate is antisymmetric in both color and flavor.
The symmetry-breaking pattern is 
$SU(2)_L\times SU(2)_R\times SU(3)_c\rightarrow SU(2)_L\times SU(2)_R\times SU(2)_c$,
in other words, the global symmetries are unschated and there are no Goldstone bosons in this phase. However, the Higgs mechanism gives rise to five massive gauge bosons, the number given by the number of broken generators. It should be mentioned that the existence of the 2SC phase is not as rigorous as that
of the CFL phase. If the couplings in the NJL model are sufficiently strong, it exists. If not, there may be a direct transition to the unpaired quark phase with no 2SC window~\cite{rajagopal1,rusterneutral,abuki}.  
There is a huge body of literature on the color superconducting phases and their properties. A number of aspects, including gauge (in)dependence of the gap, neutrality constraints and applications to compact stars, Bose-Einstein condensation of pions and kaons, inhomogeneous condensates, Meissner effect, temperature effects, and transport properties. In addition to the reviews and other papers already mentioned, relevant work can be found in e.g. Refs.~\cite{soncolor,bedak0,klevansky,browncolor,pisarskicolor,pisarskicolor2,manuelcfl,casacfl,reddycfl,bedak,baymneutral,rajagopal1,gerhold,kryj,huang,rischkeden,blacolor,rusterstar,blacolor2,krycfl,kry2cfl,rusterneutral,klim1,klim2,bubshov,ren,2sc,dobadocfl,shovkovy,buballarev,huangcolor,polcolor,gerholdcfl,abuki,andreascolor,kry3cfl,paglicfl,sedracfl,irancolor,baslercolor,partykacolor,holo0,inca,sedra2,dyson,ping,armen2,holo,baymrev,pawlow,fariasnjl,brown0,rochacfl,dirk,kojo3,grav,braun2,olek1,olek2,goscfl,geisel,khunja,alfordvisco,newbubrev}.

In this paper, we take a closer look at the two-flavor quark-meson diquark model as a low-energy description of the 2SC phase of QCD at finite baryon density. The model was studied in the context of the functional renormalization group (FRG) in Refs.~\cite{pawlow,braun2}. The starting point is the QCD Lagrangian with quark and  gluonic degrees of freedom at an initial RG scale and the value of the strong coupling constant 
$\alpha_s$ at that scale. Integrating high-energy modes in the path integral, the quark-gluon vertex generates four-quark interactions via two-gluon exchange. Combining this with (continuous) Hubbard-Stratonovich tranformations, one can access the low-energy regime that includes symmetry breaking and the formation of bound states. At low density, the scalar-pseudoscalar channel is more important, while at large density
the diquark channel is the dominant one. The corresponding couplings depend on the density as well as the temperature. 
Based on this picture, we write down the most general Lagrangian with the most relevant degrees of freedom such that it is renormalizable. This is in contrast to the NJL model which is
nonrenormalizable and whose predictions depend on a UV cutoff (see e.g. Refs.~\cite{shovkovy,buballarev} for a review, Refs.~\cite{fariasnjl,newbubrev} for work on the regularization of the NJL model, and Ref.~\cite{newbubrev} for remarks regarding renormalization of the QMD
model). Renormalizability is of course not a fundamental requirement of a useful model or theory, but it has calculational advantages: The ultraviolet divergences that appear in loop diagrams are cancelled by renormalizing the
parameters in the bare Lagrangian and the addition of higher-dimensional operators is not required.
In Ref.~\cite{patrick}, it was shown how one can use the on-shell scheme~\cite{sirlin1,sirlin2,hollik1,hollik2} to consistently determine the parameters of the quark-meson model by expressing them in terms of the meson masses $m_{\sigma}$ and $m_{\pi}$, and the pion decay constant $f_{\pi}$.  Consistently here means that the matching is done in the same approximation as the calculation of the thermodynamic potential. Ignoring this requirement, as is often done in the literature, may lead to unphysical results, such as the onset of pion condensation at $T=0$ at finite isospin chemical potential
$\mu_I\neq m_{\pi}$. Bose condensation at zero temperature is a second-order transition, which takes place exactly at the mass of the condensing boson. While the masses and couplings in the scalar-psudoscalar sector can be determined using
input from vacuum physics, this is not the case in the diquark sector. The new renormalized couplings are treated as free parameters in the model. While we do not scan the parameter space, we select a few reasonable parameter sets and show that sensible results for the phase diagram, and the speed of sound are obtained.

\section{Lagrangian}
In this section, we discuss the Lagrangian and the symmetries
of the quark-meson diquark model. As mentioned in the Introduction, the effective degrees of freedom are sigma, pions, quarks, and diquarks. The fields are denoted by $\sigma$, $\vec{\pi}$, $\psi$, and $\Delta_a$ where $a$ is a color index. The field $\psi$ is a flavor doublet and  color triplet in the fundamental representation,
\begin{eqnarray}
\psi&=&
\left(\begin{array}{c}
\psi_{ur}\\
\psi_{dr}\\
\psi_{ug}\\
\psi_{dg}\\
\psi_{ub}\\
\psi_{db}\\
\end{array}\right)\;.
\end{eqnarray}
Finally, $\psi_c$ is the flavor doublet with color $c$
\begin{eqnarray}
\psi_c&=&
\left(\begin{array}{c}
\psi_{uc}\\
\psi_{dc}\\
\end{array}\right)\;,
\end{eqnarray}
\begin{widetext}
Introducing the diquark degrees of freedom as $\Delta_a\sim \bar{\psi}_b\tau_2\epsilon_{abc}\gamma_5\psi_c^C$, the Minkowski space Lagrangian of the quark-meson diquark model can be written as
\begin{eqnarray}
\nonumber
{\cal L}&=&
{1\over2}(\partial_{\mu}\sigma)(\partial^{\mu}\sigma)
+{1\over2}(\partial_{\mu}\pi_0)(\partial^{\mu}\pi_0)
+\left(\partial_{\mu}+i\mu_I\delta_{\mu0}\right)\pi^+
\left(\partial^{\mu}-i\mu_I\delta^{\mu0}\right)\pi^-
-{1\over2}m^2(\sigma^2+\vec{\pi}^2)-{\lambda\over24}(\sigma^2+\vec{\pi}^2)^2
\\ \nonumber&&+h\sigma
+\left(\partial_{\mu}+2i{\mu}_a\delta_{\mu0}\right)
\Delta^{\dagger}_a\left(\partial^{\mu}-2i{\mu}_a\delta^{\mu0}\right)\Delta_a
-m_{\Delta}^2\Delta^{\dagger}_a\Delta_a
-{\lambda_3\over12}(\sigma^2+\vec{\pi}^2)\Delta^{\dagger}_a\Delta_a
-{\lambda_{\Delta}\over6}\left(\Delta^{\dagger}_a\Delta_a\right)^2
\\ &&
+\Bar{\psi}(i\slashed{\partial}+\gamma^0\hat{\mu})\psi-g\bar{\psi}[\sigma+i\gamma^5\vec{\pi}\cdot\vec{\tau}]\psi
+{1\over2}g_{\Delta}\Bar{\psi}^C_b\Delta_a\gamma_5\tau_2
\epsilon_{abc}\psi_c
+{1\over2}g_{\Delta}\Bar{\psi}_b\Delta_a^{\dagger}\gamma_5\tau_2\epsilon_{abc}
\psi_c^C\;,
\label{lagrangian}
\end{eqnarray}
\end{widetext}
where $\tau_a$ act in flavor space and $\epsilon_{abc}$ is the completely antisymmetric tensor in color space, and
$\psi^C_c=C\bar{\psi}_c^T$ is the charge conjugate spinor with $C=i\gamma^0\gamma^2$ being the charge conjugation matrix.
Furthermore, $\hat{\mu}$ is the quark chemical potential matrix, $\mu_I$ is the isospin chemical potential for the charged mesons, and $\mu_a$ is the chemical potential for the diquark $\Delta_a$.
In order to discuss the symmetries of the effective Lagrangian,
it is convenient to organize the meson fields in a matrix,
$\Sigma=\sigma\mathbb{1}+i\vec{\pi}\cdot\vec{\tau}$.
The matrix $\Sigma$ transforms as $L\Sigma R^{-1}$
under $SU(2)_L\times SU(2)_R$. The product $\Sigma^{\dagger}\Sigma$ is a building block for invariant terms.
Using the cyclicity of the trace $\langle A\rangle$ of a matrix $A$, it follows that $\langle\Sigma^{\dagger}\Sigma\rangle=\sigma^2+\vec{\pi}^2$ is invariant. The term $h\sigma={1\over2}h\langle\Sigma+\Sigma^{\dagger}\rangle$, on the other hand, is invariant only if $L=R$, i.e. for vector transformations. Finally, writing the interaction term $\bar{\psi}[\sigma+i\gamma^5\vec{\pi}\cdot\vec{\tau}]\psi={1\over2}\bar{\psi}_L\Sigma\psi_R+{1\over2}\bar{\psi}_R\Sigma^{\dagger}\psi_L$, it is clear that this term is also invariant. The next nontrivial term in Eq.~(\ref{lagrangian}) is the quark-diquark interaction term.
Using that $\psi_c^C=i\gamma_2\psi_c^*$, the relevant term is
\begin{eqnarray}
\nonumber\Bar{\psi}_b\gamma_5\tau_2\epsilon_{abc}\psi_c^C&=&
i\bar{\psi}_{b,L}\gamma_5\gamma_2\tau_2\epsilon_{abc}\psi^*_{c,L}\\
&&
+i\bar{\psi}_{b,R}\gamma_5\gamma_2\tau_2\epsilon_{abc}\psi^*_{c,R}\;,
\end{eqnarray}
so the left-handed and right-handed fields decouple. The first term transforms as 
\begin{eqnarray}
\label{leftterm}
i\bar{\psi}_{b,L}\gamma_5\gamma_2\tau_2\epsilon_{abc}\psi^*_{c,L}&\rightarrow&
\nonumber i\bar{\psi}_{b,L}e^{-i\tau_a\alpha_a}\gamma_5\\&&\times\gamma_2\tau_2\epsilon_{abc}e^{-i\tau_a^*\alpha_a}\psi^*_{c,L}\;.
\end{eqnarray}
Using the relation 
\begin{eqnarray}
\label{relatione}
\tau_2e^{-i\tau_a^*\alpha_a}=e^{i\tau_a\alpha_a}\tau_2\;,
\end{eqnarray}
one finds that the term in Eq.~(\ref{leftterm}) is invariant under $SU(2)_L$ transformations.
The other term is invariant under $SU(2)_R$. 
It also implies that $\Delta_a$ ($a=1,2,3$) transforms as a singlet under 
$SU(2)_L\times SU(2)_R$ and that the Lagrangian is invariant under this symmetry.

Let us next briefly consider the global $SU(3)_c$ symmetry. 
The quark-diquark interaction terms can be written as
\begin{eqnarray}
\label{gaugie}
\sim\bar{\psi}\lambda_2\psi^C\Delta_3+\bar{\psi}\lambda_5\psi^C\Delta_2    
+\bar{\psi}\lambda_7\psi^C\Delta_1
+\text{h.c}\;,    
\end{eqnarray}
where $\lambda_a$ are the Gell-Mann matrices. The quark fields transform as
\begin{eqnarray}
\psi\rightarrow e^{i\lambda_a\theta_a}\psi\;.
\end{eqnarray}
From this equation, one can work out how different terms transform under $SU(3)_c$ and in particular that Eq.~(\ref{gaugie}) is invariant. Note however that terms in Eq.~(\ref{gaugie})
are not separately invariant. One of the three $SU(2)_c$ subgroups of $SU(3)_c$ is generated by the three Gell-Mann matrices $\{\lambda_1, \lambda_2,\lambda_3\}$. Denoting the corresponding $SU(2)_c$ transformation by
$e^{i\vec{\alpha}\cdot\vec{\lambda}}$, where $\alpha_i$ ($i=1,2,3$) are parameters, the fermion fields transform as 
\begin{eqnarray}
    \psi \rightarrow e^{i\vec{\lambda}\Vec{\alpha}}\psi\;.
\end{eqnarray}
Using the relation $e^{-i\vec{\lambda}\Vec{\alpha}}\lambda_2e^{-i\vec{\lambda}^*\Vec{\alpha}} = \lambda_2$, which is the three-flavor version of Eq.~(\ref{relatione}), it is seen that the diquark field $\Delta_3$ transforms trivially under $SU(2)_c$, $\Delta_3\rightarrow\Delta_3$. However, since $\Delta_3$ is not invariant under $SU(3)_c$, the diquark condensate breaks the global $SU(3)_c$ symmetry down to $SU(2)_c$. Note also that the diquark condensate $\Delta_0$ itself is invariant under $SU(2)_L\times SU(2)_R$, which follows directly from Eq.~(\ref{leftterm}).

\section{Thermodynamic potential}
We next assume that the real field $\sigma$ and the
charged field $\Delta_3$ have nonzero real expectation values
$\phi_0$ and $\Delta_0$, i.e. we write
\begin{eqnarray}
\label{shift1}
\sigma&=&\phi_0+\tilde{\sigma}\;,\\
\Delta_3&=&\Delta_0+\tilde{\Delta}_3\;,
\label{shift2}
\end{eqnarray}
where $\tilde{\sigma}$ and $\tilde{\Delta}_3$ are fluctuating quantum fields with vanishing expectation values. The shifts Eq.~(\ref{shift1})--(\ref{shift2}) give rise to the tree-level thermodynamic potential 
\begin{eqnarray}
\nonumber
\Omega_0&=&{1\over2}m^2\phi_0^2+
(m_{\Delta}^2-4\mu_3^2)\Delta_0^2+{\lambda\over24}\phi_0^4
-h\phi_0
\\ &&
+{\lambda_3\over12}\phi_0^2\Delta_0^2+{\lambda_{\Delta}\over6}{\Delta}_0^4
\;,
\label{treepot}
\end{eqnarray}
where $\mu_3={1\over2}(\mu_{ur}+\mu_{dg})$.
We next consider loop corrections. Our approximation is to include the one-loop corrections from the quarks, but treat the mesons and diquarks at tree level. 
It is convenient to introduce the Nambu-Gorkov basis
\begin{eqnarray}
\Psi&=&
\left(\begin{array}{c}
\psi\\
\psi^C\\
\end{array}\right)\;,
\label{gorkov}
\end{eqnarray}
In this basis, the inverse quark propagator reads
\begin{eqnarray}
D^{-1}(p)&=&    
\left(\begin{array}{cc}
p\!\!\!/-g\phi_0+\gamma_0\hat{\mu}&ig_{\Delta}\tau_2\lambda_2\gamma^5\Delta_0\\
ig_{\Delta}\tau_2\lambda_2\gamma^5\Delta_0&p\!\!\!/-g\phi_0-\gamma_0\hat{\mu}\\
\end{array}\right)\;,
\end{eqnarray}
where we have replaced the tensor $\epsilon_{3ab}$ in color space with the Gell-Mann matrix $i\lambda_2$.
The spectrum is found by solving $\det D^{-1}(p)=0$ with respect to $p_0$. This yields~\cite{shovkovy}
\begin{eqnarray}
    E^\pm_{ub}(p) &=& E(p)\pm\mu_{ub}\;,\\
    E^\pm_{db}(p) &=& E(p)\pm\mu_{db}\;,\\
    E_{\Delta^{\pm}}^{\pm}(p)&=&E_{\Delta}^{\pm}(p)\pm\delta\mu\;,
\end{eqnarray}
where
\begin{eqnarray}
E(p)&=&\sqrt{p^2+g^2\phi_0^2}\;,\\
E_{\Delta}^{\pm}(p)&=&
\sqrt{(E(p)\pm\bar{\mu})^2+g_{\Delta}^2\Delta_0^2}\;,\\
\bar{\mu}&=&{1\over2}(\mu_{ur}+\mu_{dg})={1\over2}(\mu_{ug}+\mu_{dr})\;,\\
\delta\mu&=&{1\over2}(\mu_{dg}-\mu_{ur})={1\over2}(\mu_{dr}-\mu_{ug})\;.
\end{eqnarray}
In the remainder of this work we use a common quark chemical potential $\bar{\mu}$ for all colors and flavors, which implies
that $\delta\mu=0$, $\mu_3=\bar{\mu}$, and $\mu_B=3\bar{\mu}$.

The one-loop contribution to the thermodynamic potential, $\Omega_1$,
from a single fermion $f$ with dispersion relation $E(p)$ is given by 
\begin{eqnarray}
\Omega_1&=&-2\Lambda^{-2\epsilon}\int_p\left\{E(p)+T\log\left[1+e^{-\beta(E(p)\pm\mu_f)}\right]\right\}\;,   
\label{finit}
\end{eqnarray}
where $\mu_f$ is the chemical potential of the fermion, $\beta=1/T$, $\Lambda$ is the renormalization scale,
and the integral $\int_p$ is defined in Eq.~(\ref{convention}).
The integral of the first term in Eq.~(\ref{finit})
is divergent, while the integral of the second term is convergent. In the present case, the first term reads
\begin{eqnarray}
\nonumber
\Omega_1&=&
-\Lambda^{-2\epsilon}\int_p\left[
E_{ub}^{\pm}(p)+E_{db}^{\pm}(p)+2E^{\pm}_{\Delta^{\pm}}(p)
\right]\\
&=&
-4\Lambda^{-2\epsilon}\int_p\left[
E(p)+E^{+}_{\Delta}(p)+E^{-}_{\Delta}(p)
\right]\;.
\label{div}
\end{eqnarray}
The integral in Eq.~(\ref{div}) is divergent in the ultraviolet and we need to isolate the divergences.
We will use dimensional regularization, as explained in Appendix~\ref{appa}.
The first term, denoted by $\Omega_1^{a}$, is straightforward to evaluate using Eq.~(\ref{i0}),
\begin{eqnarray}
\Omega_1^{a}&=&{2g^4\phi_0^4\Lambda^{-2\epsilon}\over(4\pi)^2}
\left[{1\over\epsilon}+\log{\Lambda^2\over g^2\phi_0^2}+{3\over2}+{\cal O}(\epsilon)\right]\;.  
\label{exactie}
\end{eqnarray}
The second and third terms in Eq.~(\ref{div}), denoted by $\Omega_1^{b}$, require some extra work. We are not able to calculate the integral of $E^{\pm}_{\Delta}(p)$ directly in dimensional regularization. Instead, the strategy is to
construct a subtraction term that has the same divergent behavior as the original integrand, which can be calculated in dimensional regularization. Moreover, this term should not introduce any new artificial infrared divergences. This term is then subtracted from $E^{\pm}_{\Delta}(p)$ such that the integral converges in $d=3$ dimensions. In order not to change the thermodynamic potential, the subtraction term is added to $\Omega_1^{(b)}$ again. This strategy isolates the poles in the subtracted term, leaving us with a finite integral in three dimensions that must be evaluated numerically. Expanding the dispersion relations about $\bar{\mu}=0$ yields the divergent part
\begin{widetext}
\begin{eqnarray}
\Omega_1^{b,\rm div}&=&-8\Lambda^{-2\epsilon}\int_p\sqrt{p^2+g^2\phi_0^2+g_{\Delta}^2{\Delta}_0^2}
-4\Lambda^{-2\epsilon}\int_p{\bar{\mu}^2g_{\Delta}^2{\Delta}_0^2\over(p^2+g^2\phi_0^2+g_{\Delta}^2{\Delta}_0^2)^{3\over2}};. 
\end{eqnarray}
Using the expressions for the integrals Eqs.~(\ref{i0}) and~(\ref{i2}), we obtain
\begin{eqnarray}
\Omega_{1}^{b,\rm div}&=&{4(g^2\phi_0^2+g_{\Delta}^2{\Delta_0}^2)^2\Lambda^{-2\epsilon}\over(4\pi)^2}\left[{1\over\epsilon}+\log{\Lambda^2\over g^2\phi_0^2+g_{\Delta}^2\Delta_0^2}+{3\over2}+{\cal O}(\epsilon)\right]
-{16\bar{\mu}^2g_{\Delta}^2{\Delta}_0^2\Lambda^{-2\epsilon}\over(4\pi)^2}\left[{1\over\epsilon}
+\log{\Lambda^2\over g^2\phi_0^2+ g_{\Delta}^2\Delta_0^2}+
{\cal O}(\epsilon)\right]\;.
\label{omediv}
\end{eqnarray}
The subtracted part is defined as
\begin{eqnarray}
\Omega_{1,\rm num}^{}&=&\Omega_1^{b}-\Omega_1^{b,\rm div}
=-4\Lambda^{-2\epsilon}\int_p\left[
E^{+}_{\Delta}(p)+E^{-}_{\Delta}(p)
-2\sqrt{p^2+g^2\phi_0^2+g_{\Delta}^2\Delta_0^2}
-{\bar{\mu}^2g_{\Delta}^2\Delta_0^2\over(p^2+g^2\phi_0^2+g_{\Delta}^2\Delta^2_0)^{3\over2}}
\right]\;,   
\label{numdef}
\end{eqnarray}
which is regular in the limit $\epsilon\rightarrow0$.
Adding Eqs.~(\ref{exactie}),~(\ref{omediv}), and~(\ref{numdef}), we obtain the unrenormalized one-loop contribution
\begin{eqnarray}
\Omega_1 &=& {2g^4\phi_0^4\Lambda^{-2\epsilon}\over(4\pi)^2}\left[\frac{1}{\epsilon}+\log{\left(\frac{\Lambda^2}{g^2\phi_0^2}\right)}+\frac{3}{2}\right]+\frac{4(g^2\phi_0^2+g_\Delta^2\Delta_0^2)^2\Lambda^{-2\epsilon}}{(4\pi)^2}\left[\frac{1}{\epsilon}+\log{\left(\frac{\Lambda^2}{g^2\phi_0^2+g_\Delta^2\Delta_0^2}\right)}+\frac{3}{2}\right]\nonumber\\
\nonumber
&&-\frac{16\bar{\mu}^2g_\Delta^2\Delta_0^2\Lambda^{-2\epsilon}}{(4\pi)^2}\left[\frac{1}{\epsilon}+\log{\left(\frac{\Lambda^2}{g^2\phi_0^2+g_\Delta^2\Delta_0^2}\right)}\right]
-4\Lambda^{-2\epsilon}\int_p\left[E_\Delta^+(p)+E_\Delta^-(p)-2\sqrt{p^2+g^2\phi_0^2+g^2_{\Delta}\Delta_0^2}
\right.\\ &&\left.
-\frac{\bar{\mu}^2g^2_{\Delta}\Delta_0^2}{(p^2+g^2\phi_0^2+g^2_{\Delta}\Delta_0^2)^{3/2}}\right]\;.
\label{unrenorm}
\end{eqnarray}
Renormalization is carried out by replacing the bare masses, couplings, and fields in terms of their renormalized running counterparts. The relations between bare and renormalized quantities are
\begin{align}
\label{renorm11}
m^2&= Z_{m^2}m^2_{\ms}\;,&
\quad {\lambda}&=\Lambda^{2\epsilon}Z_{\lambda}\lambda_{\ms}\;,&
\quad \phi_0^2&=Z_{\pi}\phi_{0,\ms}^2\;,&
\quad h&= \Lambda^{-\epsilon}Z_hh_{\ms}\;,
\\
m^2_{\Delta}&=Z_{m^2_{\Delta}}m^2_{\Delta,\ms}\;,&
\quad {\lambda_3}&= \Lambda^{2\epsilon}Z_{\lambda_3}\lambda_{3,\ms}\;,&
\quad \lambda_{\Delta}&= \Lambda^{2\epsilon}Z_{\lambda_{\Delta}}\lambda_{\Delta,\ms}\;,&
\quad \Delta_0^2&=Z_{\Delta}\Delta_{0,\ms}^2
\;.
\label{renorm22}
\end{align}
where the subscript $\ms$ is a reminder that parameters are running and we are using the $\overline{\rm MS}$ scheme, as explained in Appendix~\ref{appa}. Note that the couplings are multiplied by appropriate factors of  $\Lambda^{\epsilon}$ so that the renormalized couplings have the same dimensions outside $d+1=4$.~\footnote{The renormalized Yukawa and quark-diquark
couplings $g_{\ms}(\Lambda)$ and $g_{\Delta,\ms}(\Lambda)$ are multiplied by $\Lambda^{\epsilon}$.}  
The renormalization constants in Eqs.~(\ref{renorm11})--(\ref{renorm22}) are
\begin{align}
\label{c1}
Z_{m^2}&=1+{\delta m^2_{\ms}\over m^2_{\ms}}\;,&
\quad Z_{\lambda}&=1+{\delta\lambda_{\ms}\over\lambda_{\ms}}\;,&
\quad Z_{\pi}&=1+\delta Z_{\pi}^{\ms}\;,&
\quad Z_h&=1+{\delta h_{\ms}\over h_{\ms}}\;,
\\
Z_{m^2_{\Delta}}&=1+{\delta m^2_{\Delta,\ms}\over m^2_{\Delta,\ms}}\;,&
\quad Z_{\lambda_3}&=1+{\delta\lambda_{3,\ms}\over\lambda_{3,\ms}}\;,&
\quad Z_{\lambda_{\Delta}}&=1+{\delta\lambda_{\Delta.\ms}\over\lambda_{\Delta,\ms}}\;,&
\quad Z_{\Delta}&=1+\delta Z_{\Delta,\ms}\;,
\label{c2}
\end{align}
The counterterm $\Omega_1^{\rm ct}$ follows directly from the tree-level potential Eq.~(\ref{treepot}),

\begin{eqnarray}
\nonumber
\Omega_1^{\rm ct}&=&{1\over2}\delta m^2_{\ms}\phi_{0}^2
+{1\over2}m^2_{}\phi_{0}^2\delta Z_{\pi}^{\ms}+\delta m_{\Delta,\ms}^2\Delta_{0}^2
+(m_{\Delta}^2-4\bar{\mu}^2)\Delta_{0}^2\delta Z_{\Delta}^{\ms}+{\delta\lambda_{\ms}\Lambda^{2\epsilon}\over24}\phi_{0}^4
+{\lambda \over12}\phi_{0}^4\delta Z_{\pi}^{\ms}
\\ &&
\nonumber
-{\delta h}_{\ms}\Lambda^{-\epsilon}\phi_{0}-{1\over2}h_{}
\phi_{0}\delta Z_{\pi}^{\ms}
+{\delta\lambda_{3}\Lambda^{2\epsilon}\over12}\phi_{0}^2
\Delta_{0}^2+
{\lambda_3\over12}\phi_{0}^2\Delta_{0}^2
\delta Z_{\pi}^{\ms}+{\lambda_3\over12}\phi_{0}^2
\Delta_{0}^2\delta Z_{\Delta}^{\ms}
\\ &&
+{\delta\lambda_{\Delta,\ms}\Lambda^{2\epsilon}\over6}\Delta_{0}^4+
{\lambda_{\Delta}\over3}\Delta_{0}^4\delta Z_{\Delta}^{\ms}
\;.
\label{count}
\end{eqnarray}
The counterterms in\ the $\overline{\rm MS}$-scheme in Eq.~(\ref{count}) are determined by requiring that they cancel exactly against the poles in $\epsilon$ in Eq.~(\ref{unrenorm}). We note that this does not determine all the  counterterms separately, in some cases it is only various combinations that are determined in this manner. For example, the absence of a $\phi_0^2$-term in Eq.~(\ref{unrenorm}) implies that the sum
$\delta m^2_{\ms}+m^2\delta Z_{\pi}^{\ms}$ vanishes, but
individually the counterterms must be determined in a different way. Moreover, the above procedure does not determine the 
counterterms $\delta g^2_{\ms}$ and $\delta g_{\Delta,\ms}^2$ in
the expressions~\footnote{In our approximation, $Z_{\psi}=1$.}
\begin{eqnarray}
Z_{g^2}=1+{\delta g^2_{\ms}\over g^2_{\ms}}\;,
\hspace{2cm}
Z_{g_{\Delta}^2}=1+{\delta g^2_{\Delta,\ms}\over g^2_{\Delta,\ms}}\;.
\end{eqnarray}
The counterterms related to the vacuum physics
are determined in Appendix~\ref{fixing}, where we discuss
the relation between the on-shell and $\overline{\rm MS}$ schemes and the determination of the running parameters in terms of physical masses and the pion decay constant.
They are given in Eqs.~(\ref{dm})--(\ref{dzpi}).
The remaining counterterms in Eq.~(\ref{count}) are determined in Appendix~\ref{quarkprop} using the method just described.
They are given in Eqs.~(\ref{new1})--(\ref{newl}).
Adding Eqs.~(\ref{treepot}),~(\ref{unrenorm}), and~(\ref{count}), we obtain the renormalized one-loop thermodynamic potential expressed in terms of running masses, couplings, and fields as
\begin{eqnarray}
\nonumber
\Omega_{0+1}&=&{1\over2}m^2_{\ms}\phi_{0,\ms}^2
+\left(m^2_{\Delta,\ms}-4\bar{\mu}^2\right)\Delta_{0,\ms}^2+{\lambda_{\ms}\over24}\phi_{0,\ms}^4
-h_{\ms}\phi_{0,\ms}
+{\lambda_{3,\ms}\over12}\phi_{0,\ms}^2\Delta^2_{0,\ms}
+{\lambda_{\Delta,\ms}\over6}\Delta_{0,\ms}^4
\\&&
\nonumber
+{2g_{\ms}^4\phi_{0,\ms}^4\over(4\pi)^2}\left[\log{\Lambda^2\over g_{\ms}^2\phi_{0,\ms}^2}+{3\over2}\right]+
{4(g_{\ms}^2\phi_{0,\ms}^2+g_{\Delta,\ms}^2\Delta_{\ms}^2)^2\over(4\pi)^2}\left[\log{\Lambda^2\over g_{\ms}^2\phi_{0,\ms}^2+g_{\Delta,\ms}^2\Delta_{0,\ms}^2}+{3\over2}\right]
\\ &&
-\frac{16\bar{\mu}^2g_{\Delta,\ms}^2\Delta_{0,\ms}^2}{(4\pi)^2}\log{\frac{\Lambda^2}{g^2_{\ms}\phi_{0,\ms}^2+g_{\Delta,\ms}^2\Delta_{0,\ms}^2}}
+\Omega_{1,\rm num}\;.
\label{nesten}
\end{eqnarray}
The running masses, couplings, and fields satisfy a set of renormalization group equations whose solutions can also be found in Appendices~\ref{fixing} and~\ref{quarkprop}.
Substituting the solutions to the renormalization group equations 
(\ref{sol1})--(\ref{sol4}) and (\ref{sol5})--(\ref{sol9})
for the running parameters into Eq.~(\ref{nesten}), we obtain
the final expression for the thermodynamic potential $\Omega_{0+1}$
\begin{eqnarray}
\nonumber
\Omega_{0+1}&=&
{3\over4}m_{\pi}^2f_{\pi}^2\left\{1-{12m_q^2\over(4\pi)^2f_{\pi}^2}
m_{\pi}^2F^{\prime}(m_{\pi}^2)\right\}{\phi_0^2\over f_{\pi}^2}+{2m_q^4\over(4\pi)^2}\left({9\over2}+\log{\frac{m_q^2}{g_0^2\phi_0^2}} +2\log{\frac{m_q^2}{g_0^2\phi_0^2+g_{\Delta,0}^2\Delta_{0}^2}}\right)\frac{\phi_0^4}{f_\pi^4}\\
&&
\nonumber
-{1\over4}m_{\sigma}^2f_{\pi}^2\left\{1+{12m_q^2\over(4\pi)^2f_{\pi}^2}
\left[\left(1-{4m_q^2\over m_{\sigma}^2}\right)F(m_{\sigma}^2)+{4m_q^2\over m_{\sigma}^2}
-F(m_{\pi}^2)-m_{\pi}^2F^{\prime}(m_{\pi}^2)
\right]
\right\}{\phi_0^2\over f_{\pi}^2}
\\ &&
\nonumber
+{1\over8}m_{\sigma}^2f_{\pi}^2\left\{
1+{12m_q^2\over(4\pi)^2f_{\pi}^2}\left[
\left(1-{4m_q^2\over m_{\sigma}^2}\right)F(m_{\sigma}^2)
-F(m_{\pi}^2)-m_{\pi}^2F^{\prime}(m_{\pi}^2)
\right]
\right\}{\phi_0^4\over f_{\pi}^4}
\\ &&
\nonumber
-{1\over8}m_{\pi}^2f_{\pi}^2\left\{1-{12m_q^2\over(4\pi)^2f_{\pi}^2}
m_{\pi}^2F^{\prime}(m_{\pi}^2)\right\}{\phi_0^4\over f_{\pi}^4}
-m_{\pi}^2f_{\pi}^2\left\{1-{12m_q^2\over(4\pi)^2f_{\pi}^2}m_{\pi}^2F^{\prime}(m_{\pi}^2)\right\}{\phi_0 \over f_{\pi}}
\\ && 
\nonumber
+\left\{m_{\Delta,0}^2-4\bar{\mu}^2\left[1+{4g_{\Delta,0}^2\over(4\pi)^2}\left(\log{m_q^2\over g_0^2\phi_0^2+g_{\Delta,0}^2\Delta_0^2}-F(m_\pi^2)-m_\pi^2F'(m_\pi^2)\right)\right]\right\}\Delta_0^2
\\ &&
\nonumber
+{\lambda_{3,0}\over12}\phi_0^2\Delta_0^2 +{\lambda_{\Delta,0}\over6}\Delta_0^4 + \frac{12g_0^2g_{\Delta,0}^2}{(4\pi)^2}\phi_0^2\Delta_0^2+\frac{6g_{\Delta,0}^4}{(4\pi)^2}\Delta_0^4
\\ &&
\nonumber
+{4g_{\Delta,0}^4\over (4\pi)^2}\left\{\log{m_q^2\over g_0^2\phi_0^2+g_{\Delta,0}^2\Delta_0^2}-F(m_\pi^2)-m_\pi^2F'(m_\pi^2)\right\}\Delta_0^4
\\
&& 
\nonumber
+{4g_0^2g_{\Delta,0}^2\over (4\pi)^2}\left\{\log{m_q^2\over g_0^2\phi_0^2+g_{\Delta,0}^2\Delta_0^2}-F(m_\pi^2)-m_\pi^2F'(m_\pi^2)\right\}\phi_0^2\Delta_0^2+\Omega_{1,\rm num}
+\Omega^{\bar{\mu},T}
\;,
\label{final}
\end{eqnarray}
where the last term is
\begin{eqnarray}
   \Omega^{\bar{\mu},T} 
&=&-4T
\int_p\bigg\{\log\left[1+e^{-\beta(E\pm\bar{\mu}_{})}\right]
+2\log\left[1+e^{-\beta(E^{\pm}_{\Delta})}\right]\bigg\}\;.
\label{improved}
\end{eqnarray}
Note that thermodynamic potential in Eq.~(\ref{improved})
is independent of the scale $\Lambda$, as a consequence of
the renormalization group improvement.
\end{widetext}

In order to find the zero-temperature behavior of $c_s^2$ at large $\bar{\mu}$, we consider the thermodynamic potential $\Omega_{0+1}$
at $T=0$. If we are deep into the 2SC phase,
we have $\phi_0\ll\Delta_0$.
The thermodynamical potential in Eq.~(\ref{final}) then reduces to
\begin{eqnarray}
\nonumber
\Omega_{0+1}&=&
(m_{\Delta,0}^2-4\bar{\mu}^2)\Delta_0^2+{\lambda_{\Delta,0}\over6}\Delta_0^4
-{8\over(4\pi)^2}\bar{\mu}^4
\\ && 
\nonumber
-{16g_{\Delta,0}^2\over(4\pi)^2}\log{m_q^2\over g_{\Delta,0}^2\Delta_0^2}\bar{\mu}^2\Delta_0^2
\nonumber
\\ &&
+{4g_{\Delta,0}^4\over (4\pi)^2}\left[\log{m_q^2\over g_{\Delta,0}^2\Delta_0^2}+{3\over2}\right]\Delta_0^4\;.
\end{eqnarray}
In the same approximation, the gap equation reads
\begin{eqnarray}
\nonumber
0&=&m_{\Delta,0}^2-4\bar{\mu}^2
+{\lambda_{\Delta,0}\over3}\Delta_0^2
-{16g_{\Delta,0}^2\over(4\pi)^2}\left(\log{m_q^2\over g_{\Delta,0}^2\Delta_0^2}-1\right)\bar{\mu}^2
\\ &&
+{8g_{\Delta,0}^4\over (4\pi)^2}\left[\log{m_q^2\over g_{\Delta,0}^2\Delta_0^2}+1\right]\Delta_0^2\;.
\label{eqn_gap_delta}
\end{eqnarray}
Assuming $\bar{\mu}\gg\Delta_0$, we can solve the gap equation explicitly.
Denoting the gap by $\bar{\Delta}_0$ in this regime, we find
\begin{eqnarray}
\bar{\Delta}_0^2&=&{m_q^2\over g_{\Delta,0}^2}e^{{(4\pi)^2\over4g_{\Delta,0}^2}-1}\;. 
\end{eqnarray}
In other words, $\Delta_0$ approaches a constant $\bar{\Delta}_0$
as $\bar{\mu}\rightarrow\infty$.
This is in contrast to the behavior of the gap at tree-level. In this
case, the gap grows linearly with $\bar{\mu}$. The difference is caused
by the logarithms introduced by adding quark loops.

The pressure $p$ is given by minus the thermodynamic potential, where
$\Omega_{0+1}$ is evaluated at the solution to the gap equation, i.e. 
$p(\bar{\mu})=-\Omega_{0+1}(\bar{\mu},\bar{\Delta}_0)$.
At high densities, the pressure is 
\begin{eqnarray}
p&=&
{8\over(4\pi)^2}\bar{\mu}^4+{16g_{\Delta,0}^2\over(4\pi)^2}\bar{\mu}^2\bar{\Delta}_0^2\;.
\end{eqnarray}
The quark density $n_q$ is
\begin{eqnarray}
n_q&=&-{\partial\Omega_{0+1}\over\partial\bar{\mu}}\;.    
\end{eqnarray}
The energy density $\epsilon$ is given by the Legendre transform
\begin{eqnarray}
\epsilon&=&-p+\bar{\mu}n_q\;,    
\end{eqnarray}
where the right-hand side is again evaluated $\Delta_0=\bar{\Delta}_0$.
This yields
\begin{eqnarray}
\epsilon&=&
{24\over(4\pi)^2}\bar{\mu}^4+{16g_{\Delta,0}^2\over(4\pi)^2}\bar{\mu}^2\bar{\Delta}_0^2\;.    
\end{eqnarray}
The speed of sound squared is
\begin{eqnarray}
c_s^2&=&{dp\over d\epsilon}={{dp\over d\bar{\mu}}\over {d\epsilon\over d\bar{\mu}}}\;.
\end{eqnarray}
In the large-$\bar{\mu}$ limit, this reduces to
\begin{eqnarray}
    c_s^2 &=& 
\frac{\bar{\mu}^2+g_{\Delta,0}^2\Bar{\Delta}_0^2}{3\bar{\mu}^2+g_{\Delta,0}^2\Bar{\Delta}_0^2}
\approx{1\over3}\left(1+{2\over3}{g_{\Delta,0}^2\bar{\Delta}_0^2\over\bar{\mu}^2}\right)\;.   
    \label{speed}
    \end{eqnarray}
In this form it is easy to see that the speed of sound approaches the conformal limit from above.

\section{Results and discussion}
\label{results}
In this section, we present and discuss our numerical results. 
The chiral condensate $\phi_0$ and the superconducting gap $\Delta_0$ are determined by finding the global minimum of \(\Omega\). This means, solving the equations
\begin{eqnarray}
{\partial\Omega_{}\over\partial\phi_0}=0\;,
\hspace{2cm}
{\partial\Omega_{}\over\partial\Delta_0}=0\;.
\end{eqnarray}
We start by analyzing the tree-level thermodynamic potential Eq.~(\ref{treepot}). Although there are no finite-temperature effects, the fact 
that we can easily calculate the condensate and the gap analytically gives us some insight. 
In the remainder, we work in the chiral limit and the minimization conditions are
\begin{eqnarray}
\label{phi01}
m^2\phi_0+{\lambda\over6}\phi_0^3
+{\lambda_3\over6}\phi_0\Delta_0^2&=&0\;,\\ 
(m_{\Delta}^2-4\bar{\mu}^2)\Delta_0+{\lambda_3\over6}\phi_0^2\Delta_0
+{2\over3}\lambda_{\Delta}\Delta_0^3
&=&0\;.
\label{gap2}
\end{eqnarray}
In the limiting case, where $\lambda_3=0$, i.e. when there is no
coupling between the two equations~(\ref{phi01}) and~(\ref{gap2}), we find
\begin{eqnarray}
\phi_0=\sqrt{-6m^2\over\lambda}\;,\hspace{8mm}
\Delta_0=\sqrt{3(4\bar{\mu}^2-m_{\Delta}^2)\over2\lambda_{\Delta}}\;.
\end{eqnarray}
In this case, the onset of color superconductivity takes places
at $\bar{\mu}={1\over2}m_{\Delta}$, while $\phi_0$
takes it vacuum value $f_{\pi}$ for all values of $\bar{\mu}$.
This case is unnatural since it requires the fine tuning of
a coupling constant and will not be considered in the remainder of the article.
In the case where $\lambda_3\neq0$, we find that either $\phi_0$ or $\Delta_0=0$. Again the transition takes
place at $\bar{\mu}={1\over2}m_{\Delta}$.

We next consider the full thermodynamic potential given Eq.~(\ref{improved}). For numerical work, we use the values 
\begin{eqnarray}
m_{\pi}&=&0\;,\\
m_{\sigma}&=&500\, {\rm MeV}\;,\\
f_{\pi}&=&93\, {\rm MeV}\;.
\end{eqnarray}
Moreover, the quark mass in the vacuum is
\begin{eqnarray}
m_q&=&300\, {\rm MeV}\;.    
\end{eqnarray}
This gives the numerical values
\begin{eqnarray}
\label{r1}
m^2&=&-(487\,{\rm MeV})^2\;,\\
\label{r2}
\lambda&=&65.8\;,\\
g&=&3.2\;,\\
h&=&0\;.
\end{eqnarray}
These parameters should be compared to their tree-level counterparts, given
by Eq.~(\ref{tre1})--(\ref{treq}), whose numerical values are
\begin{eqnarray}
\label{t1}
m^2&=&-(353.6\,{\rm MeV})^2
\;,\\
\label{t2}
\lambda&=&86.7
\;,\\
g&=&3.2\;,\\
h&=&0\;.
\end{eqnarray}
Comparing Eqs.~(\ref{r1})--(\ref{r2}) with Eqs.~(\ref{t1})--(\ref{t2}), we see that
there are substantial renormalization effects. The parameters $g$ and $h$ are not renormalized. For the coupling $g$ it is the result of choosing the renormalization scale $\Lambda_0$
according to Eq.~(\ref{refscale}), cf. Eq.~(\ref{rung}).
For the coupling $h$, it is a consequence of the Goldstone theorem being respected when loop corrections are taken correctly into account, cf. Eq.~(\ref{hrun}).

\begin{figure}
\centering
\includegraphics[width=8.5cm]{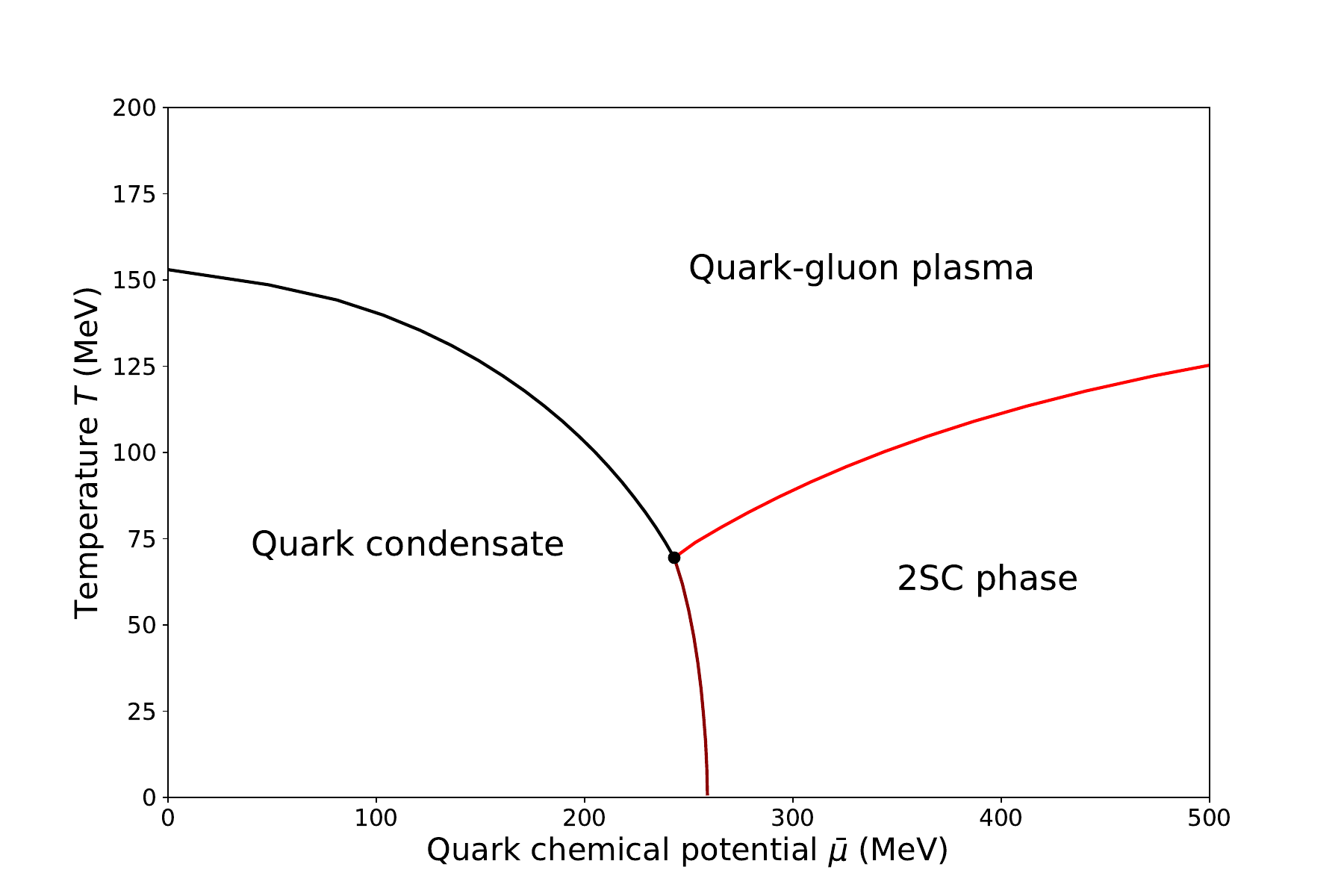}
\\
\includegraphics[width=8.5cm]{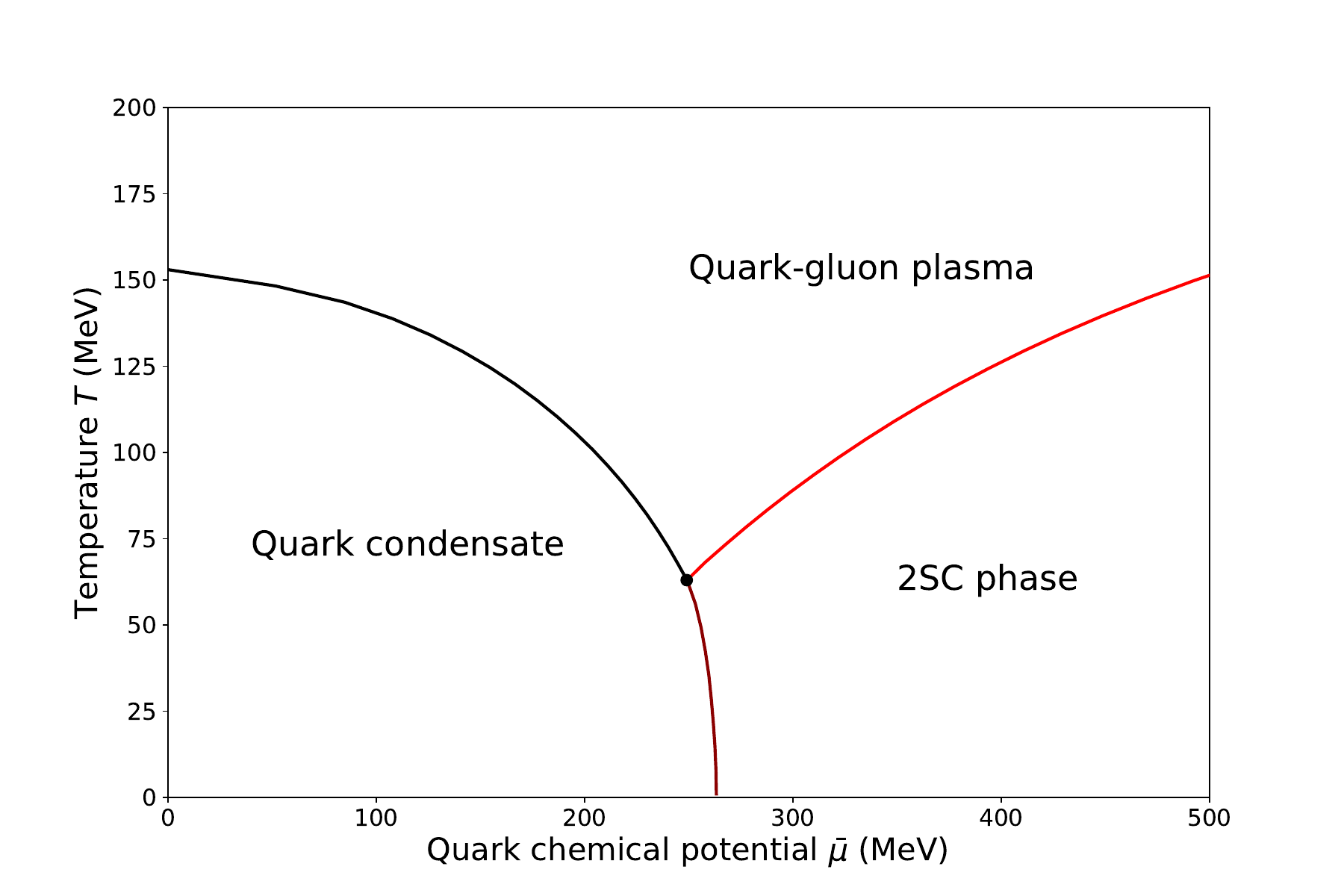}
\caption{Four sample phase diagrams in the $\bar{\mu}$--$T$ plane. See main text for details.}
\label{mut}
\end{figure}

The remaining parameters are $m_{\Delta,0}^2$,
$\lambda_{3,0}$, $\lambda_{\Delta,0}$, and $g_{\Delta,0}^2$, which cannot be related to any vacuum physics. As explained, we consider them free parameters, but we
can make arguments for reasonable values. We have seen that the transition at $T=0$ takes place
at $\bar{\mu}={1\over2}m_{\Delta}$  at tree-level. Values of $m_{\Delta}$ in the 500-1000 MeV range therefore seem reasonable.  In the NJL model, the diquark coupling $G_{\Delta}$ can be obtained from the
coupling $G$ in the standard Lagrangian through a Fierz transformation giving the result $G_{\Delta}={3\over4}G$, for a derivation see \cite{buballarev}, appendix A. Since this gives rise to superconducting gaps of approximately 200 MeV, they allowed it to vary between zero (obviously giving no 2SC phase) and $G_{\Delta}=G$.
The couplings $g$ and $g_{\Delta}$ play a similar role in the quark-meson-diquark model and we vary $g_{\Delta}$ around $g$. Finally, a simple but sensible choice is  $\lambda_3$ and $\lambda_{\Delta}$ to be of the order of $\lambda$.
In this paper, we are not scanning the parameter space, but choose a few values according to the discussion above. Specifically, we choose the four parameter sets
\begin{align}
{\color{red}
{\rm Set\,1}:}\,m_{\Delta,0}=500\,{\rm MeV}\;,
g_{\Delta,0}=2g\;,\\
{\color{blue}
{\rm Set\,2}:}\,m_{\Delta,0}=500\,{\rm MeV}\;,
g_{\Delta,0}={3\over2}g\,,\\
{\color{green!50!black}
{\rm Set\,3}:}\,m_{\Delta,0}=600\,{\rm MeV}\;,
g_{\Delta,0}=2g\,,\\
{\color{orange}
{\rm Set\,4}:}\,m_{\Delta,0}=600\,{\rm MeV}\;,
g_{\Delta,0}={3\over2}g\,,
\end{align}
with \(\lambda_{\Delta,0} = \lambda_0/4\) and \(\lambda_{3,0} = \lambda_0\) for all sets to have a symmetric potential. The result is a few sample phase diagrams shown in 
the $\bar{\mu}$--$T$ plane in Fig.~\ref{mut}.
They show three phases, namely the chirally broken phase with a quark condensate, the 2SC phase, and the quark-gluon plasma phase, where both condensates vanish. The point where all three phases meet is a tricritical point. As is evident from the panels, there is some flexibility in the model. The position of the tricritical point moves around, as do the phase boundaries. Qualitatively, our phase diagram is in good agreement with NJL model calculations, which gives us some confidence in the quark-meson diquark model.
In both models, it is necessary to introduce new terms in the Lagrangian in order to describe the color superconducting phase. The couplings of these operators cannot be determined by matching the model to known vacuum physics. It may be possible to determine
the parameters in the QMD model using properties of observed neutron stars such as masses, radii, and tidal deformabilities. Using the QMD model to describe the 2SC phase in hybrid stars, these macroscopic properties depend on the unknown couplings. A Bayesian inference approach can then be used to determine the couplings as done recently in Refs.~\cite{bayes,bayes2} in the context of an NJL model description of hadronic matter and quark matter, respectively.

\begin{figure}[htb!]
\includegraphics[width=8.5cm]{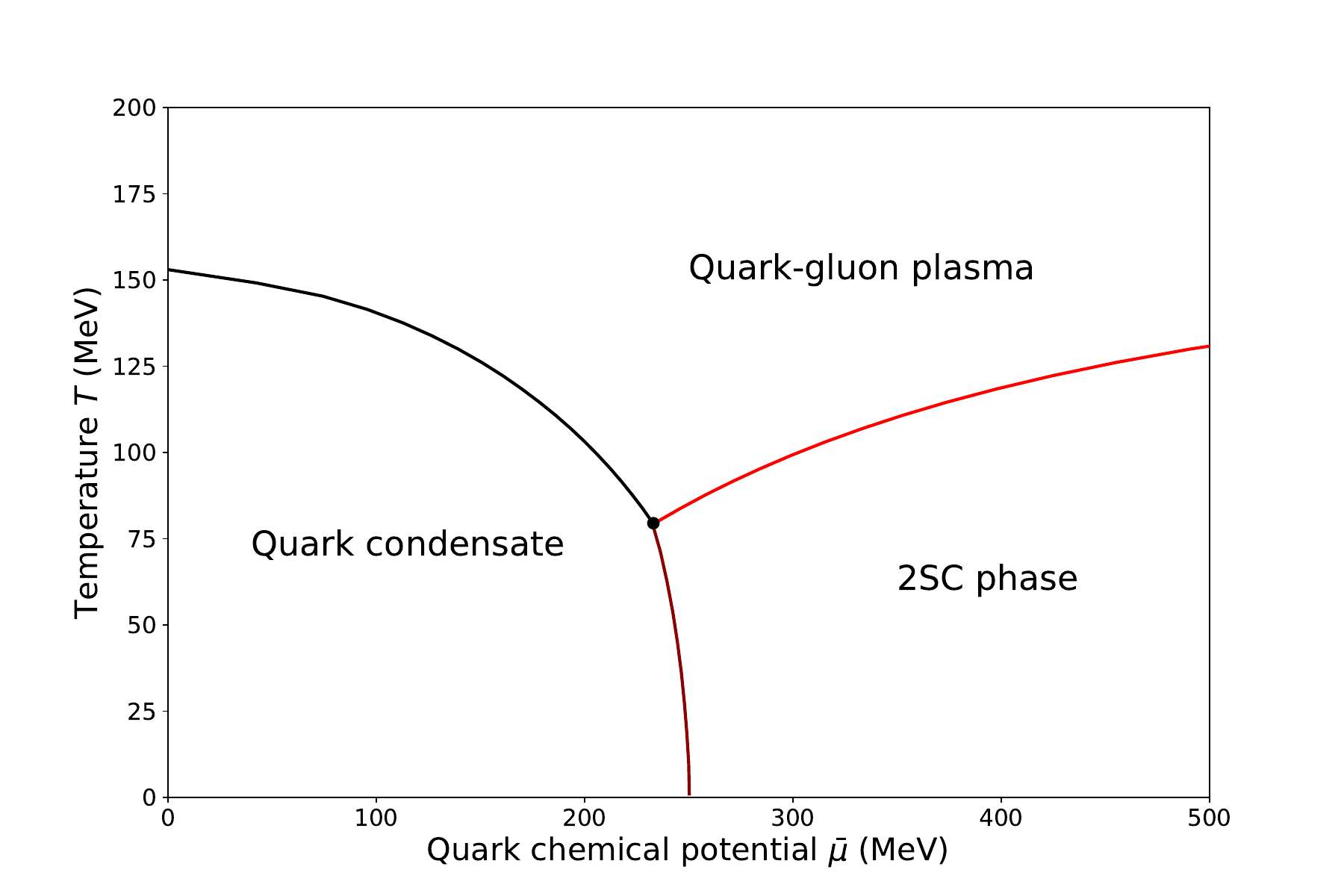}
\\
\includegraphics[width=8.5cm]{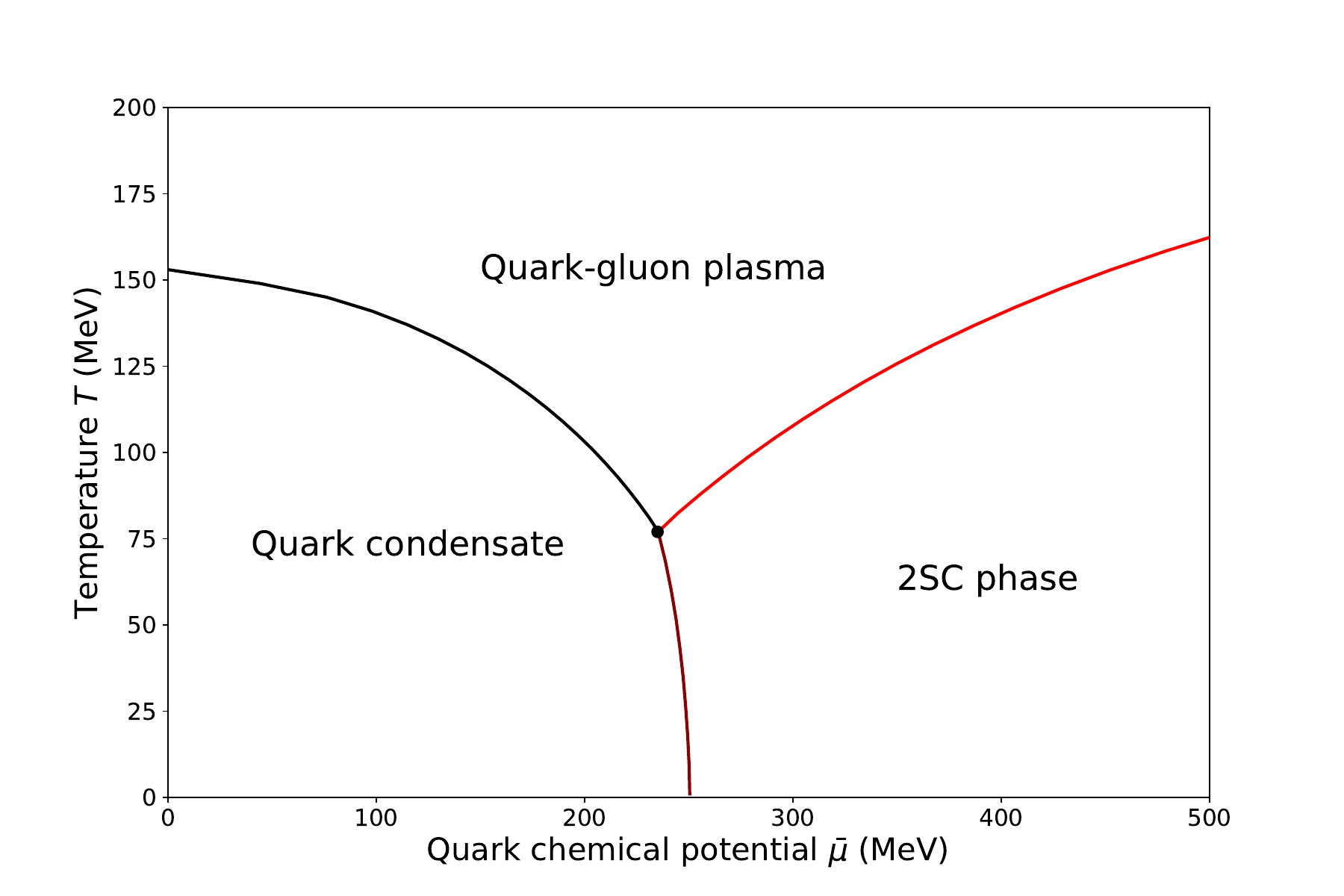}
\end{figure}

\begin{figure}[htb!]
\includegraphics[width=8.5cm]{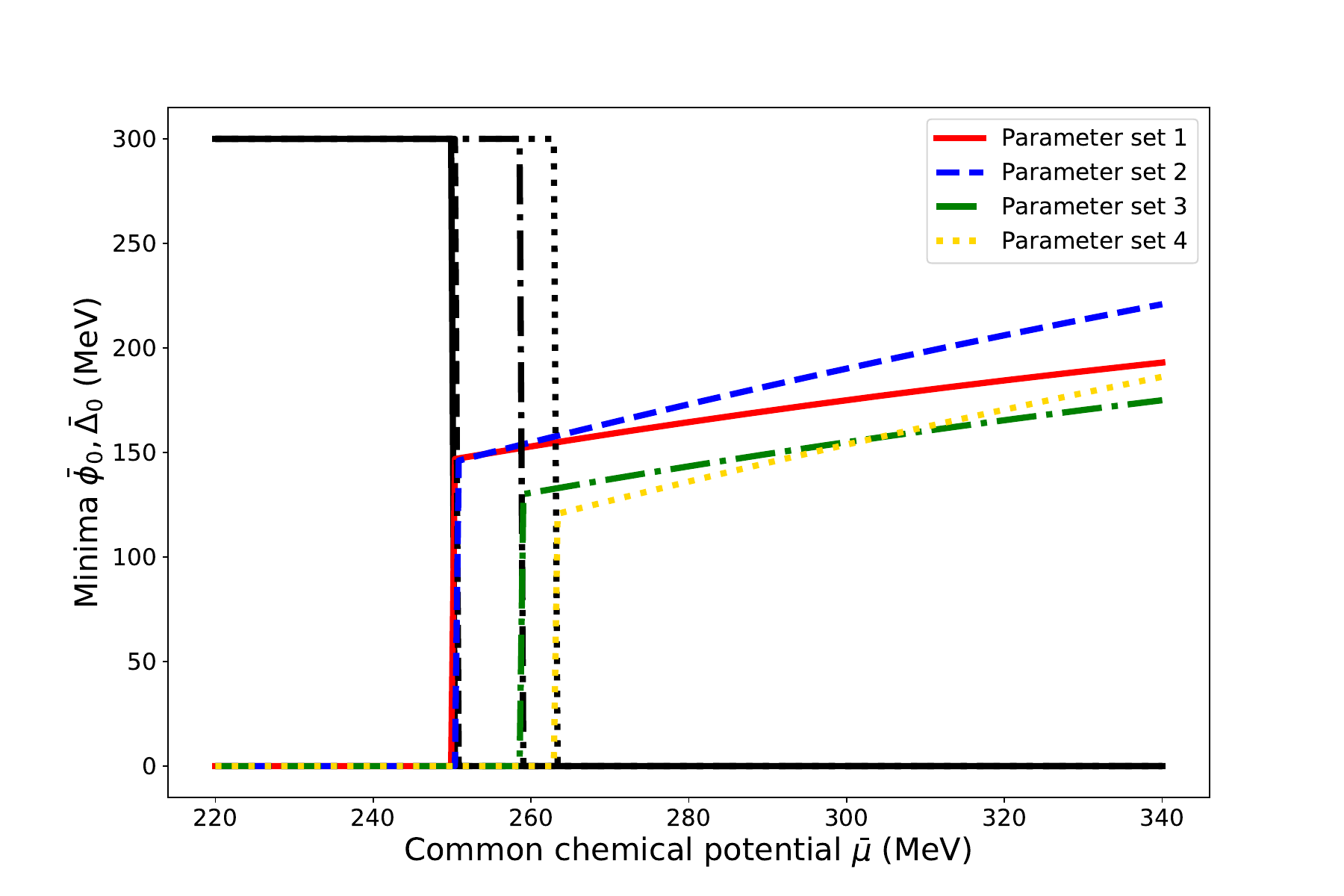}
\caption{Condensates $\phi_0$ and $\Delta_0$ as functions of the quark chemical potential $\bar{\mu}$ for $T=0$. See main text for details.}%
\label{condensates}
\end{figure}

In Fig.~\ref{condensates}, we show the zero-temperature condensates $\phi_0$ and $\Delta_0$ as functions of $\bar{\mu}$. The transition is taking place at $\bar{\mu}\approx280$ MeV and is first order. As discussed above, the gap saturates as $\bar{\mu}\rightarrow\infty$. This asymptotic behavior is first seen for much larger values of $\bar{\mu}$
than shown in Fig.~\ref{condensates}. Note also that the condensate $\phi_0$ takes on its vacuum value
all the way up to the transition. This is also the case for the thermodynamic quantities, which is the Silver-Blaze property. Consequently, the speed of sound vanishes in this phase, cf. Fig.~\ref{speedo} below.

We next discuss the speed of sound $c_s$, which is very important for compact stars as it measures the stiffness of the EoS. A detailed discussion of $c_s$ at finite isospin chemical potential $\mu_I$ and finite baryon chemical potential $\mu_B$ was recently given in Ref.~\cite{roma}. In Fig.~\ref{speedo}, we show the zero-temperature result for the speed of sound $c_s$ as a function of $\bar{\mu}$ for the four different parameter sets. The black dashed horizontal curve indicates the conformal limit $c_s=1/\sqrt{3}$. In all four cases, 
$c_s$ has a maximum after which it approaches the conformal limit from above. This result is at variance with perturbative calculations. Perturbative calculations, which assume a trivial ground state, i.e. no color superconducting gap, suggest that
the conformal limit is approached from below.
In a recent paper~\cite{geisel}, the authors include the effects of a superconducting gap in their perturbative calculations at next-to-leading order. They work in the chiral limit and further assume that the gap is small compared to the chemical potential. Again, the results indicate an approach to the conformal limit from below.
However, baryon chemical potentials considered here are not necessarily accessible in pQCD since the series may not be convergent. Moreover, results for the speed of sound reported in Ref.~\cite{fuji} using hard-dense-loop resummed perturbation theory, indicate that $c_s$ approaches the conformal limit from above.
We cannot compare our results for the speed of sound with lattice simulations due to the sign problem. However, there is another system of interest where Monte Carlo methods can be used, namely QCD at finite isospin density. In Ref.~\cite{endrodieos}, the authors calculate $c_s$ as a function of $\mu_I$ at $T=0$ for physical pion masses $m_{\pi}$. The speed of sound squared increases from zero at $\mu_I=m_{\pi}$ i.e. at onset of pion condensation to approximately $c_s^2=0.56$ at
$\mu_I/m_{\pi}=0.77$ (with a small dependence on the lattice). After this peak, $c_s^2$ decreases. Similar results have been obtained in Refs.~\cite{isoabbott0,isoabbott}, where the results extends to larger values of the isospin chemical potential $\mu_I$.
The results seem to indicate that $c_s$ approaches the conformal limit, qualitatively in the same fashion as found in this work.
In Ref.~\cite{isokojo}, the authors consider the quark-meson model at finite isospin using the same techniques as in the present paper. The behavior of the speed of sound is in very good agreement with the lattice results over a large range of chemical potentials, in particular
that $c_s$ peaks at some finite value of $\mu_I$ before it approaches the conformal limit from above.~\footnote{The pressure, energy density, and gap equations have the same structure, so these results seem to be generic for these type of low-energy models.}

\begin{figure}[htb!]
\includegraphics[width=8.5cm]{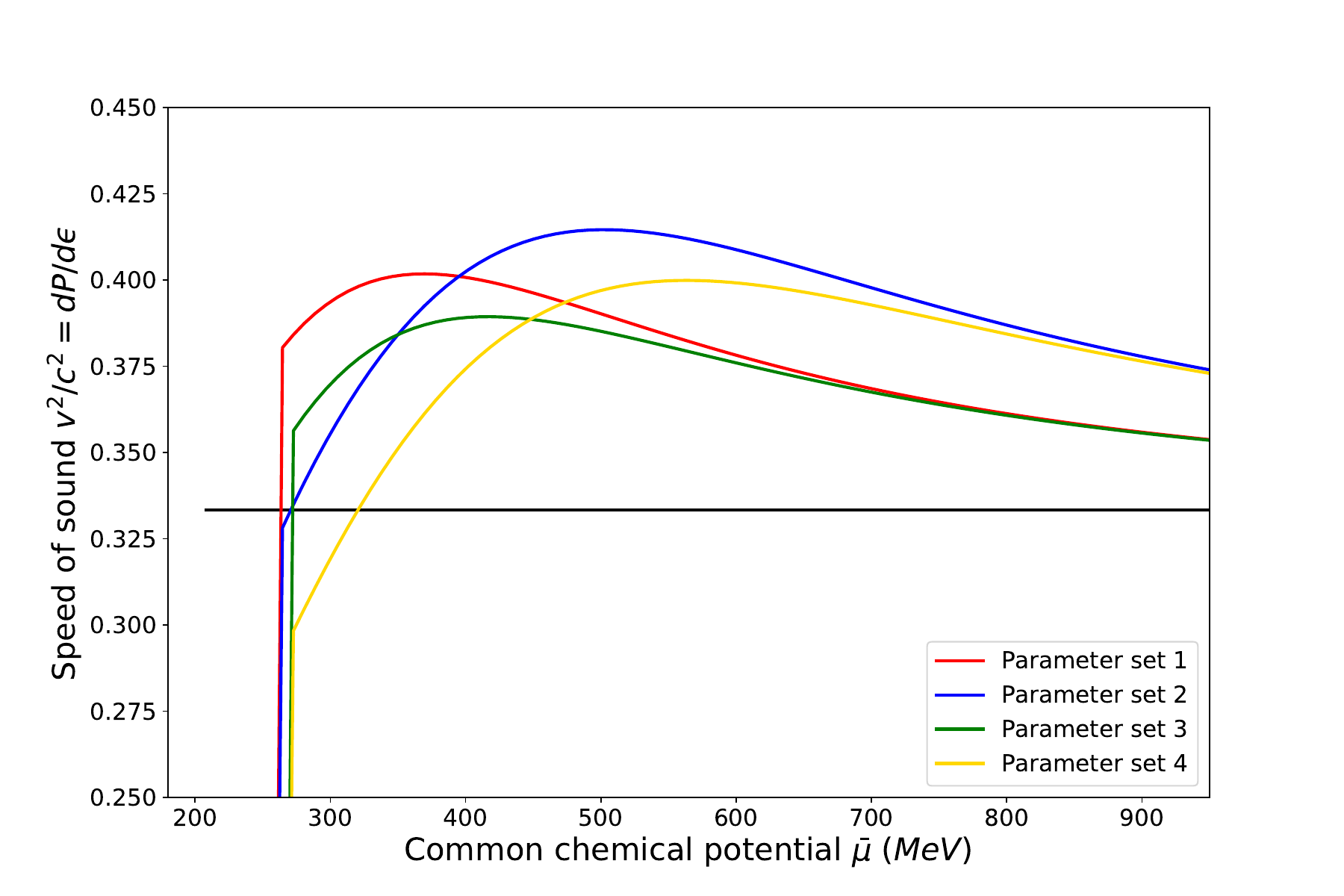}
\caption{Speed of sound  squared $c_s^2$ 
as a function of baryon chemical potential $\bar{\mu}$
for $T=0$.
See main text for details.}%
\label{speedo}
\end{figure}

In the present paper, we have renormalized the thermodynamic potential in a mean-field approximation that includes fermionic fluctuations. We have derived a set of renormalization
group equations for the running masses couplings, and fields. 
The solutions to these renormalization group equations have been used to improve the thermodynamic potential. The resulting thermodynamic potential has been used to map out the phase diagram in the $\bar{\mu}$--$T$ plane. We have also computed the speed of sound $c_s$ and found that
it fairly quickly approaches the conformal limit $c_s={1\over\sqrt{3}}$ from above, which is a generic property of the model.
Since this paper is a first exploratory study, there are several directions for further work.
It would be of interest to consider the competition between a diquark condensate and an inhomogeneous chiral condensate as done in the NJL model calculations of Ref.~\cite{dirk}.
Similarly, it would be of interest to include the possibility of pion condensation at finite $\mu_I$ and map out the phase diagram in the $\mu_I$--$\mu_B$ plane.
Such calculations should be combined with imposing compact star conditions, i.e. electric charge and color charge neutrality. In the 2SC phase, the chemical potentials can be expressed in terms of three independent chemical potentials, namely $\mu_B$, $\mu_e$, and $\mu_8$, where $\mu_e$ is the electron chemical  potential and $\mu_8$ is the chemical potential related to to color charge $Q_8$.
Two neutrality constraints then leave us with a single independent chemical potential. Work in these directions is in progress~\cite{nyttpapir}.

\appendix

\section{Integrals needed}
\label{appa}
In the calculations, we encounter a few ultraviolet divergent loop integrals 
in three dimensions that need to be regularized. We use dimensional regularization in $d$ dimensions. The integrals are defined as 
\begin{eqnarray}
\int_p&=&\left({e^{\gamma_{E}}\Lambda^2\over4\pi}\right)^{\epsilon}
\int{d^dp\over(2\pi)^d}\;,
\label{convention}
\end{eqnarray}
where $\Lambda$ is the renormalization scale associated with the $\overline{\rm MS}$
scheme. The specific integrals needed in $d=3-2\epsilon$ dimensions are
\begin{eqnarray}
\label{i0}
\int_p\sqrt{p^2+m^2}&=&-{m^4\over2(4\pi)^2}\left({\Lambda\over m}\right)^{2\epsilon}
\left[{1\over\epsilon}+{3\over2}+{\cal O}(\epsilon)\right]
\;,\\
\int_p{1\over(p^2+m^2)^{3\over2}}&=&{4\over(4\pi)^2}\left({\Lambda\over m}\right)^{2\epsilon}
\left[{1\over\epsilon}+{\cal O}(\epsilon)\right]\;.
\label{i2}
\end{eqnarray}
We also need some four-dimensional vacuum integrals in Minkowski space. The specific integrals needed 
in $d=4-2\epsilon$ dimensions are 
\begin{eqnarray}
\nonumber
A(m^2)&=&\int_p{1\over p^2-m^2}={im^2\over(4\pi)^2}\left({\Lambda\over m}\right)^{2\epsilon}\left[{1\over\epsilon}+1+{\cal O}(\epsilon)\right]\;,    \\ 
\label{adef}
&& \\
\nonumber
B(p^2)&=&\int_q{1\over[q^2-m^2][(p+q)^2-m^2]}\\
&=&
\label{bdef}
{i\over(4\pi)^2}\left({\Lambda\over m}\right)^{2\epsilon}\left[{1\over\epsilon}+F(p^2)
+{\cal O}(\epsilon)\right]\;,
\\
B^{\prime}(p^2)&=&{i\over(4\pi)^2}F^{\prime}(p^2)\;,
\end{eqnarray}
where $F(p^2)=2-2q\arctan({1\over q})$ with $q=\sqrt{{4m^2\over p^2}-1}$.

\section{Parameter fixing in the vacuum}
\label{fixing}
In this Appendix, we discuss the parameter fixing of the quark-meson part of the model. There are four parameters in the Lagrangian, namely $m^2$, $\lambda$, $g$, and $h$. In the vacuum, the tree-level field-dependent masses are
\begin{eqnarray}
\label{rela1}
m_{\pi}^2&=&m^2+{\lambda\over6}\phi_0^2\;,\\
\label{rela2}
m_{\sigma}^2&=&m^2+{\lambda\over2}\phi_0^2\;.
\end{eqnarray}
We identify the pion decay constant $f_{\pi}$ with the value of $\phi_0$ that minimizes the thermodynamic potential in the vacuum, i.e. $f_{\pi}$ satisfies the equation
${\partial \Omega\over\partial\phi_0}|_{\phi_0=f_{\pi}}=0$.
Evaluating these expression at the minimum $f_{\pi}$ of the thermodynamic potential yields the physical pole masses. Solving Eqs.~(\ref{rela1})--(\ref{rela2}) with respect to $m^2$ and $\lambda$ yields
\begin{eqnarray}
\label{tre1}
m^2&=&-{1\over2}(m_{\sigma}^2-3m_{\pi}^2)\;,\\
\lambda&=&3{(m_{\sigma}^2-m_{\pi}^2)\over f_{\pi}^2}\;.
\label{tre2}
\end{eqnarray}
The parameter $h$ is determined by the equation of motion, ${\partial \Omega_0\over\partial\phi_0}|_{\phi_0=f_{\pi}}=m^2f_{\pi}+{\lambda\over6}f_{\pi}^3-h=0$, or
\begin{eqnarray}
h&=&m_{\pi}^2f_{\pi}\;.    
\label{treh}
\end{eqnarray}
Finally, the quark mass is given by 
\begin{eqnarray}
m_q&=&gf_{\pi}\;,    
\label{treq}
\end{eqnarray}
which follows from the scalar-quark term in Eq.~(\ref{lagrangian}).

In the on-shell scheme~\cite{sirlin1,sirlin2,hollik1,hollik2}, the idea is that the tree-level relations are satisfied order by order in the loop expansion. Thus the renormalized parameters in the on-shell scheme are equal to their tree-level counterparts and they do not run with the renormalization scale. In order to achieve this, one must choose the counterterms appropriately, meaning that they exactly cancel the loop corrections to the relevant $n$-point functions.

\begin{figure}[htb!]
        \centering
        \begin{tikzpicture}
            \begin{feynman}
                \vertex (a);
                \vertex [left=of a] (b) {\(\sigma\)};
                \vertex [above right=of a] (c);
                \vertex [below right=of a] (d);
                \diagram* {
                  (b) --  (a) --  [quarter left] (c) -- [fermion, half left] (d) -- [quarter left] (a),
                  };
            \end{feynman};\node[left=2cm]{};
        \end{tikzpicture}
        \caption{Tadpole diagram for the $\sigma$ field.}
        \label{taddy}
    \end{figure}
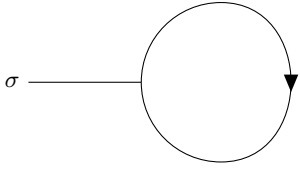

We first consider the one-point function $\Gamma_{\sigma}^{(1)}$ for the sigma field. The one-loop contribution is shown in Fig.~\ref{taddy}, where a solid line is a sigma propagator
and a solid line with an arrow is a quark propagator.
    
\begin{figure}[htb!]
    \begin{tikzpicture}
        \begin{feynman}
            \vertex (s);
            \vertex [left =of s] (a) ;
            \vertex [right =of s] (b);
            \vertex [right =of b] (c);
            \diagram* {
            (a) -- [solid] (s) -- [fermion, half left, edge label=] (b) -- [fermion, half left, edge label=] (s);
            (b) -- [solid] (c);
            };
        \end{feynman}
        \node[left=2cm]{\(i\Sigma^{\rm 1PI}_{\sigma}(p^2) = \)};
    \end{tikzpicture}
\caption{1PI self-energy diagram for $\sigma$.}
\label{1pi}
    \end{figure}
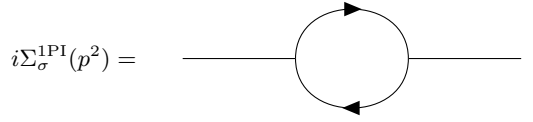

\begin{figure}[htb!]
        \begin{tikzpicture}
            \begin{feynman}
                \vertex (s);
                \vertex [left=of s] (a) {};
                \vertex [right=of s] (b) {};
                \vertex [above=of s] (c);
                \vertex [above left=of c] (d);
                \vertex [above right=of c] (e);
                \diagram* {
                  (a) -- [solid] (s)  -- [solid] (b),
                  (s)  --  [edge label =] (c) ,
                  (c) --[quarter left, edge label=] (d),
                  (d) --[fermion, half left, edge label=] (e),
                  (e) --[quarter left, edge label=] (c),
                  };
            \end{feynman};
    \node[left=2cm]{\(i\Sigma^{\rm tadpole}_{\sigma}= \)};
        \end{tikzpicture}  
    \caption{Tadpole self-energy diagrams for $\sigma$.} 
    \label{tadpole00}
\end{figure}
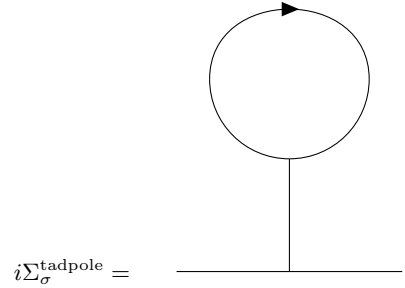

The expression for the 
one-point function including the loop correction is
\begin{eqnarray}
\Gamma^{(1)}&=&m^2\phi_0-h-24ig^2\Lambda^{-2\epsilon}\phi_0A(m_q^2)-\delta\Gamma^{(1)}\;,    
\end{eqnarray}
where the integral $A(m_q^2)$ is defined in Eq.~(\ref{adef})
and $\delta\Gamma^{(1)}$ denotes the counterterm. 
Note that there is a factor of $N_c=3$ for each fermion loop.
Finding the minimum of the thermodynamic potential as function of $\phi_0$ is equivalent 
to solving the equation $\Gamma^{(1)}=0$. Normally the loop correction gives rise to a shift of the minimum, which is finite after renormalization. In order to ensure that the minimum $\phi_0$ is still located at $f_{\pi}$, the counterterm is chosen such that it exactly cancels the loop correction, i.e. 
the counterterm must be
\begin{eqnarray}
\label{tadpole}
\delta\Gamma^{(1)}&=&-24ig^2\Lambda^{-3\epsilon}
f_{\pi}A(m_q^2)\;,
\end{eqnarray}
where we have used $\phi_0\Lambda^{\epsilon}=f_{\pi}$.
We next consider the two-point function or the inverse propagator for $\sigma$ and $\vec{\pi}$, which is given by
\begin{eqnarray}
 \Gamma^{(2)}_{\sigma,\pi}(p^2)&=&p^2-m_{\sigma,\pi}^2-\Sigma_{\sigma,\pi}(p^2)+\Sigma^{\rm ct}_{\sigma,\pi}(p^2)\;,   
\end{eqnarray}
where the subscript indicates the meson in question,
$\Sigma_{\sigma,\pi}(p^2)$ is the self-energy, and 
$\Sigma^{\rm ct}_{\sigma,\pi}(p^2)$ are the counterterms.
In the on-shell scheme, the pole of the propagator (or the zero of the two-point function)
is located at $p^2=m_{\sigma,\pi}^2$, with residue one. 
These renormalization conditions can be written as
\begin{eqnarray}
\Sigma_{\sigma,\pi}(m^2_{\sigma,\pi})-\Sigma^{\rm ct}(m_{\sigma,\pi}^2)&=&0\;,\\
\Sigma_{\sigma,\pi}^{\prime}(m^2_{\sigma,\pi})-\Sigma^{\prime\rm ct}(m^2_{\sigma,\pi})&=&0\;,
\end{eqnarray}
where the prime denotes differentiation with respect to $p^2$.
The counterterms are written as
\begin{eqnarray}
\Sigma_{\sigma,\pi}^{\rm ct}(p^2)&=&-\delta m_{\sigma,\pi}^2
+(p^2-m_{\sigma,\pi}^2)\delta Z_{\sigma,\pi}
+\Sigma_{\sigma,\pi}^{\rm ct,tadpole}\;.
\end{eqnarray}
This yields
\begin{eqnarray}
\label{renorm1}
\delta m^2_{\sigma,\pi}&=&-\Sigma_{\sigma,\pi}^{\rm 1PI}(m^2_{\sigma,\pi})\;,\\
\delta Z_{\sigma,\pi}&=&\Sigma^{\prime}(m_{\sigma,\pi}^2)\;,
\label{renorm2}
\\
\Sigma_{\sigma,\pi}^{\rm ct,tadpole}&=&
\Sigma_{\sigma,\pi}^{\rm tadpole}(m_{\sigma,\pi}^2)\;,
\label{renorm3}
\end{eqnarray}
where the superscript 1PI means that only the one-particle irreducible diagrams to the self-energy are included. 
For the $\sigma$, this diagram is shown in Fig.~\ref{1pi}. 
The 1PI part of the pion and sigma self-energies are given by
\begin{eqnarray}
\Sigma_{\pi}^{\rm 1PI}(p^2)&=&-24ig^2\Lambda^{-2\epsilon}[A(m_q^2)-{1\over2}p^2B(p^2)]\;,
\label{piself}\\
\Sigma_{\sigma}^{\rm1PI}(p^2)&=&-24ig^2\Lambda^{-2\epsilon}[A(m_q^2)-{1\over2}(p^2-4m_q^2)B(p^2)]\;.
\nonumber
\\
&&
\;.
\label{sigself}
\end{eqnarray}
The tadpole diagram for $\sigma$ is shown in Fig.~\ref{tadpole00},
but we do not need the expression for it.    
Using Eqs.~(\ref{renorm1})--(\ref{renorm2}), we then obtain the counterterms
\begin{eqnarray}
\label{counter1}
\delta m_{\pi}^2&=&24ig^2\Lambda^{-2\epsilon}[A(m_q^2)-{1\over2}m^2_{\pi}B(m^2_{\pi})]\;,\\
\nonumber
\delta m^2_{\sigma}&=&24ig^2\Lambda^{-2\epsilon}[A(m_q^2)-{1\over2}(m_{\sigma}^2-4m_q^2)B(m^2_{\sigma})]\;,\\
&&\\
\delta Z_{\pi}&=&12ig^2\Lambda^{-2\epsilon}\left[B(m_{\pi}^2)+m_{\pi}^2B^{\prime}(m^2_{\pi})\right]\;,\\
\label{counter4}
\delta Z_{\sigma}&=&12ig^2\Lambda^{-2\epsilon}\left[B(m^2_{\sigma})+(m_{\sigma}^2-4m_q^2)B^{\prime}(m^2_{\sigma})\right]\;,
\end{eqnarray}
In our approximation, there are no loop correction to the quark propagator. The counterterms are therefore
\begin{eqnarray}
\delta m^2&=&-{1\over2}(\delta m_{\sigma}^2-3\delta m_{\pi}^2)\;,\\
\delta\lambda&=&3{(\delta m_{\sigma}^2-\delta m_{\pi}^2)\over f_{\pi}^2}-\lambda{\delta f_{\pi}^2\over f_{\pi}^2}\;,\\
\delta m_q&=&0\;.
\label{dmq}
\end{eqnarray}
Eq.~(\ref{dmq}) implies that $\delta g^2=-g^2{\delta f_{\pi}^2\over f_{\pi}^2}$.
Moreover since there is no vertex correction, $\delta g^2=-g^2\delta Z_{\pi}$, which yields
\begin{eqnarray}
\label{relasjon}
\delta f_{\pi}^2&=&f_{\pi}^2\delta Z_{\pi}
\;.    
\end{eqnarray}
We finally need to determine the counterterm $\delta h$.
The one-point function can be written as $\Gamma^{(1)}=h-m_{\pi}^2f_{\pi}$, which yields 
\begin{eqnarray}
\delta h&=&[\delta m_{\pi}^2f_{\pi}+m_{\pi}^2\delta f_{\pi}]\Lambda^{-\epsilon}
+i\delta\Gamma^{(1)}\;.
\end{eqnarray}
Using Eqs.~(\ref{tadpole}), (\ref{counter1}), (\ref{counter4}), and (\ref{relasjon}), we obtain
\begin{eqnarray}  
\delta h&=&-6ig^2\Lambda^{-2\epsilon}m_\pi^2f_{\pi}\left[B(m_{\pi}^2)-m_{\pi}^2B^{\prime}(m_{\pi}^2)\right]\;.
\end{eqnarray}
After having determined the counterterms in the on-shell scheme, we
would like to relate the renormalized parameters in the $\overline{\rm MS}$-scheme to the renormalized parameters in the
$\rm{OS}$-scheme. Since the bare parameters in the two schemes
are independent of the renormalization scheme, one can 
immediately write down relations between their renormalized counterparts. 
This yields
\begin{eqnarray}
m^2_{\ms}&=&m^2+\delta m^2_{}-\delta m^2_{\ms}\;,\\
\lambda_{\ms}&=&\lambda+\delta\lambda_{}-\delta\lambda_{\ms}\;,\\
g^2_{\ms}&=&g^2+\delta g^2_{}-\delta g^2_{\ms}\;,\\
h_{\ms}&=&h+\delta h_{}-\delta h^2_{\ms}\;.
\end{eqnarray}
The counterterms in the $\overline{\rm MS}$-scheme exactly cancel 
poles in $\epsilon$ arising from the divergent loop integrals. This yields
\begin{widetext}

\begin{eqnarray}
\label{ms1}
m_{\ms}^2&=&-{1\over2}(m_{\sigma}^2-3m_{\pi}^2)\left[1+{12m_q^2\over(4\pi)^2f_{\pi}^2}\log{\Lambda^2\over m_q^2}\right]
-{12m_q^2\over(4\pi)^2f_{\pi}^2}\left[2m_q^2+{1\over2}(m_{\sigma}^2-4m_q^2)F(m_{\pi}^2)-{3\over2}m_{\pi}^2F^{\prime}(m_{\pi}^2)\right]\;,
\\
\lambda_{\ms}&=&3{m_{\sigma}^2-m_{\pi}^2\over f_{\pi}^2}
\left\{1+{12m_q^2\over(4\pi)^2f_{\pi}^2}\left[2\log{\Lambda^2\over m_q^2}+
F(m_{\pi}^2)+m_{\pi}^2F^{\prime}(m_{\pi}^2)
\right]\right\}
+{36m_q^2\over(4\pi)^2f_{\pi}^4}\left[
m_{\sigma}^2F(m_{\sigma}^2)-m_{\pi}^2F(m_{\pi}^2)
\right. \nonumber\\ &&\left.
-4m_q^2\log{\Lambda^2\over m_q^2}
-4m_q^2F(m_{\sigma}^2)
\right]\;,
\\
\label{rung}
g_{\ms}^2&=&{m_q^2\over f_{\pi}^2}\left\{1+{12m_q^2\over(4\pi)^2f_{\pi}^2}\left[\log{\Lambda^2\over m_q^2}+F(m_{\pi}^2)+m_{\pi}^2F^{\prime}(m_{\pi}^2)\right]\right\}\;,\\
\label{hrun}
h_{\ms}&=&m_{\pi}^2f_{\pi}\left\{
1+{6m_q^2\over(4\pi)^2f_{\pi}^2}\left[\log{\Lambda^2\over m_q^2}+
F(m_{\pi}^2)-m_{\pi}^2F^{\prime}(m_{\pi}^2)\right]
\right\}\;,\\
\phi_{\ms}^2&=&\left\{1-{12m_q^2\over(4\pi)^2f_{\pi}^2}
\left[\log{\Lambda^2\over m_q^2}
+F(m_{\pi}^2)+m_{\pi}^2F^{\prime}(m_{\pi}^2)
\right]\right\}\phi_0^2\;,
\label{mslast}
\end{eqnarray}    
\end{widetext}
where the counterterms are
\begin{eqnarray}
\label{dm}
\delta m^2_{\ms}&=&{12m^2g^2\Lambda^{-2\epsilon}\over(4\pi)^2}{1\over\epsilon}\;,\\
\delta\lambda_{\ms}&=&{24g^2\left[\lambda-6g^2\right]\Lambda^{-4\epsilon}\over(4\pi)^2}{1\over\epsilon}\;,\\
\delta g^2_{\ms}&=&{12g^4\Lambda^{-4\epsilon}\over(4\pi)^2}{1\over\epsilon}\;,
\\
\delta h_{\ms}&=&{6hg^2\Lambda^{-\epsilon}\over(4\pi)^2}{1\over\epsilon}\;,
\label{dh} \\
\delta Z_\pi^{\ms}&=&
-{12g^2\Lambda^{-2\epsilon}\over(4\pi)^2}{1\over\epsilon}\;.
\label{dzpi}
\end{eqnarray}
In Eqs.~(\ref{ms1})--(\ref{mslast}), we have eliminated the parameters $m^2$, $\lambda$, 
$g^2$, and $h$ in favor of the physical quantities $m_{\sigma}$, $m_{\pi}$, $f_{\pi}$, and $m_q$ using Eqs.~(\ref{tre1})--(\ref{treq}).
Taking the derivative of the running parameters 
Eqs.~(\ref{ms1})--(\ref{mslast}) with respect to the renormalization scale $\Lambda$, we obtain the renormalization group equations
\begin{eqnarray}
    \Lambda\frac{dm_{\ms}^2(\Lambda)}{d{\Lambda}} &=& \frac{24m_{\ms}^2(\Lambda)g^2_{\ms}(\Lambda)}{(4\pi)^2}\;,\\
  \Lambda  \frac{d\lambda_{\ms}(\Lambda)}{d{\Lambda}} &=& \frac{48[g_{\ms}^2(\Lambda)\lambda_{\ms}(\Lambda)-6g_{\ms}^4(\Lambda)]}{(4\pi)^2}\;,\\
 \Lambda   \frac{dg_{\ms}^2(\Lambda)}{d{\Lambda}} &=& \frac{24g_{\ms}^4(\Lambda)}{(4\pi)^2}\;,\\
    \Lambda\frac{dh_{\ms}(\Lambda)}{d{\Lambda}} &=& \frac{12g^2_{\ms}(\Lambda)h_{\ms}(\Lambda)}{(4\pi)^2}\;,    \\
 \Lambda   \frac{d\phi_{0,\ms}^2(\Lambda)}{d{\Lambda}} &=& -\frac{24\phi_{0,\ms}^2g_{\ms}^2(\Lambda)}{(4\pi)^2}\;.
 \end{eqnarray}
The solutions to these equations are
\begin{eqnarray}
\label{sol1}
m^2_{\ms}(\Lambda)&=&{m_0^2\over{1-{12g^2_{0}\over(4\pi)^2}\log{\Lambda^2\over\Lambda_0^2}}}\;, \\
\lambda_{\ms}(\Lambda)&=&{\lambda_0-{144g_0^2\over(4\pi)^2}\log{\Lambda^2\over\Lambda_0^2}\over{\Big(1-{12g_0^2\over(4\pi)^2}\log{\Lambda^2\over\Lambda_0^2}}\Big)^2}\;, \\
g^2_{\ms}(\Lambda)&=&{g_0^2\over{1-{12g_0^2\over(4\pi)^2}\log{\Lambda^2\over\Lambda_0^2}}}\;, \\
h_{\ms}(\Lambda)&=&{h_0\over{}
\sqrt{1-{12g_0^2\over(4\pi)^2}\log{\Lambda^2\over\Lambda_0^2}}}\;,
\label{sol4}
\\
\phi^2_{0,\ms}(\Lambda)&=&\left[1-{12g_0^2\over(4\pi)^2}\log{\Lambda^2\over\Lambda_0^2}
\right]\phi_0^2\;,
\end{eqnarray}
where $\Lambda_0$ is a reference scale and the constants $m_0^2$, $\lambda_0$, $g_0^2$, 
$h_0$, and $\phi_0^2$ are the values of the running parameters at the scale $\Lambda_0$.
The reference scale can be chosen at will. A convenient choice is that it satisfies the equation
\begin{eqnarray}
\label{refscale}
\log{\Lambda_0^2\over m_q^2}+
F(m_{\pi}^2)+m_{\pi}^2F^{\prime}(m_{\pi}^2)&=&0\;.
\end{eqnarray}
Note that the product $g^2_{\ms}(\Lambda)\phi_{0,\ms}^2(\Lambda)=g_{0}^2\phi_0^2$, i.e. it is independent of the renormalization scale.

\section{Remaining parameters}
\label{quarkprop}
In this Appendix, we determine the remaining counterterms $\delta m_{\Delta}^2$,
$\delta \lambda_{3}$, $\delta \lambda_{\Delta}$, and $\delta Z_{\Delta}$.
There are two ways of doing this. One way is to calculate the self-energy of the
diquark field $\Delta_a$ and determine the counterterms by requiring that the two-point function be finite. The second method is, as mentioned above, to equate the different divergent terms in the unrenormalized thermodynamic potential $\Omega$ with the corresponding counterterms. We choose the latter. Equating the divergent terms proportional to $\Delta_0^2$
$\bar{\mu}^2\Delta_0^2$, $\phi_0^2\Delta_0^2$, and 
$\Delta_0^4$ in Eqs.~(\ref{unrenorm}) and~(\ref{count}), we obtain the coupled equations
\begin{eqnarray}
\delta m_{\Delta,\ms}^2+m_{\Delta}^2\delta Z_{\Delta}^{\ms}&=&0\;,\\
-4\delta Z_{\Delta}^{\ms}-{16g_{\Delta}^2\Lambda^{-2\epsilon}\over(4\pi)^2\epsilon}&=&0\;,\\
\nonumber
{\delta\lambda_{3,\ms}\Lambda^{2\epsilon}\over12}+
{\lambda_{3}\over12}\delta Z_{\pi}^{\ms}
+{\lambda_{3}\over12}\delta Z_{\Delta}^{\ms}
\\ +{8g^2g_{\Delta}^2\Lambda^{-2\epsilon}\over(4\pi)^2\epsilon}&=&0
\\
{\delta\lambda_{\Delta,\ms}\Lambda^{-2\epsilon}\over6}+
{\lambda_{\Delta}\over3}\delta Z_{\Delta}^{\ms}+
{4g_{\Delta}^4\Lambda^{-2\epsilon}\over(4\pi)^2\epsilon}&=&0
\;.
\end{eqnarray}
Since there is no wavefunction renormalization of the quark fields or the quark-diquark
vertex, we find $\delta {g^2_{\Delta,\ms}}=-g_{\Delta}^2\Lambda^{-2\epsilon}\delta Z_{\Delta}^{\ms}$.
Using the expression for $\delta Z_{\pi}^{\ms}$ in Eq.(\ref{dzpi}), we find
\begin{eqnarray}
\label{new1}
\delta m_{\Delta,\ms}^2&=&
{4m_{\Delta}^2g_{\Delta}^2\Lambda^{-2\epsilon}\over(4\pi)^2}{1\over\epsilon}\;,\\
\delta\lambda_{3,\ms}&=&
{4\left[3\lambda_{3}g_{}^2+\lambda_{3}g_{\Delta}^2-24g_{}^2g_{\Delta}^2\right]\Lambda^{-4\epsilon}\over(4\pi)^2}{1\over\epsilon}
\;,\\
\delta\lambda_{\Delta,\ms}&=&
{8g_{\Delta}^2\left[\lambda_{\Delta}-3g_{\Delta}^2\right]\Lambda^{-4\epsilon}\over(4\pi)^2}{1\over\epsilon}
\;,\\
\delta {g^2_{\Delta,\ms}}&=&{4g_{\Delta}^4\Lambda^{-4\epsilon}\over(4\pi)^2}{1\over\epsilon}\;,\\
\delta Z_{\Delta}^{\ms}&=&-{4g_{\Delta}^2\Lambda^{-2\epsilon}\over(4\pi)^2}{1\over\epsilon}\;.
\label{newl}
\end{eqnarray}
The bare parameters are given by the sum of the renormalized running parameter and the counterterm,
for example $m_{\Delta}^2=m^2_{\Delta,\ms}+\delta m_{\Delta,\ms}^2$. The bare quantities are independent of the renormalization scale, which is used to derive the renormalization group equations for the running mass and couplings. In the limit $\epsilon\rightarrow0$, they read
\begin{eqnarray}
\label{run1}
\Lambda{dm_{\Delta,\ms}^2\over d\Lambda}&=&{8m_{\Delta,\ms}^2g_{\Delta,\ms}^2\over(4\pi)^2}\;,\\
\Lambda{d\lambda_{3,\ms}\over d\Lambda}&=&
{8(3\lambda_{3,\ms}g_{\ms}^2+\lambda_{3,\ms} g_{\Delta,\ms}^2-24g_{\ms}^2g_{\Delta,\ms}^2)\over(4\pi)^2}
\;,\\
\Lambda{d\lambda_{\Delta,\ms}\over d\Lambda}&=&
{16g_{\Delta,\ms}^2[\lambda_{\Delta,\ms}-3g_{\Delta,\ms}^2]\over(4\pi)^2}\;,\\
\Lambda{d{g_{\Delta,\ms}^2}\over d\Lambda}&=&
{8g_{\Delta,\ms}^4\over(4\pi)^2}\;,\\
\Lambda{d{\Delta_{0,\ms}^2}\over d\Lambda}&=&
-{8\Delta_{0,\ms}^2g_{\Delta,\ms}^2\over(4\pi)^2}\;.
\label{runl}
\end{eqnarray}
The solutions to Eqs.~(\ref{run1})--(\ref{runl}) are
\begin{eqnarray}
\label{sol5}
m_{\Delta,\ms}^2(\Lambda)&=&{m_{\Delta,0}^2\over1-{4g_{\Delta,0}^2\over(4\pi)^2}\log{\Lambda^2\over\Lambda_0^2}}\;,  \\
\lambda_{3,\ms}(\Lambda)&=&{\lambda_{3,0}-
{96g_0^2g_{\Delta,0}^2\over(4\pi)^2}\log{\Lambda^2\over\Lambda_0^2}
\over\left[1-{12g_0^2\over(4\pi)^2}\log{\Lambda^2\over\Lambda_0^2}\right]\left[
1-{4g_{\Delta,0}^2\over(4\pi)^2}\log{\Lambda^2\over\Lambda_0^2}\right]}\;,\\
\lambda_{\Delta,\ms}(\Lambda)&=&{\lambda_{\Delta,0}-
24{g_{\Delta,0}^4\over(4\pi)^2}\log{\Lambda^2\over\Lambda_0^2}
\over\left[1-{4g_{\Delta,0}^2\over(4\pi)^2}\log{\Lambda^2\over\Lambda_0^2}\right]^2}\;,
\\
g_{\Delta,\ms}^2(\Lambda)&=&
{g_{\Delta,0}^2\over1-{4g_{\Delta,0}^2\over(4\pi)^2}\log{\Lambda^2\over\Lambda_0^2}}\;,
\\
\Delta_{0,\ms}^2(\Lambda)&=&
\left[{1-{4g_{\Delta,0}^2\over(4\pi)^2}\log{\Lambda^2\over\Lambda_0^2}}\right]\Delta_0^2\;.
\label{sol9}
\end{eqnarray}
where $\Lambda_0$ again is the reference scale satisfying Eq.~(\ref{refscale})
and $m_{\Delta,0}^2$, $\lambda_{3,0}$, $\lambda_{\Delta,0}$, $g_{0}^2$, $g_{\Delta,0}^2$, and $\Delta_0^2$ are the values of the running parameters at that scale.
Note that the product $g^2_{\Delta,\ms}(\Lambda)\Delta_{\ms}^2(\Lambda)=g_{\Delta,0}^2\Delta_0^2$, i.e. it is independent of the renormalization scale.

\bibliographystyle{apsrmp4-1}

\end{document}